\documentclass[final,times]{elsarticle}
\usepackage[margin=0.9in]{geometry}
\usepackage{graphicx}
\usepackage{amssymb}
\usepackage{amsmath}
\usepackage{lineno}
\usepackage{placeins}
\usepackage{xcolor}
\usepackage{url} 
\usepackage{hyperref}
\journal{}

\graphicspath{{Figures/}}

\makeatletter
\def\ps@pprintTitle{%
 \let\@oddhead\@empty
 \let\@evenhead\@empty
 \def\@oddfoot{\centerline{\thepage}}%
 \let\@evenfoot\@oddfoot}
\makeatother

\begin{document}
\begin{frontmatter}

\title{Effect of new jet substructure measurements on Pythia8 tunes}

\author[label1]{Deepak Kar}
\author[label2]{Pratixan Sarmah}

\address[label1]{School of Physics, 
University of Witwatersrand\\
Johannesburg, South Africa.\\
Email: deepak.kar@cern.ch}

\address[label2]{Department of Physics, 
BITS Pilani\\
Rajasthan, India.\\
Email: pratixan123@gmail.com}

\begin{abstract}

This masters project used the recent ATLAS jet substructure measurements to see if any improvements can be made to the commonly used Pythia8 Monash and A14 tunes.

\end{abstract}


\end{frontmatter}


\section{Introduction}
\label{sec:intro}

The commonly used Pythia8~\cite{Sjostrand:2007gs,Sjostrand:2014zea} tunes, Monash~\cite{Skands:2014pea} and A14~\cite{ATL-PHYS-PUB-2014-021} are rather dated, and the latter was observed to have some tension with LEP measurements, primarily due to its lower Final State Radiation (FSR) $\alpha_{s}$ value. In last couple of years, a plethora of jet substructure~\cite{Altheimer:2012mn,Altheimer:2013yza,Marzani:2019hun,Asquith:2018igt} measurements have been published by both ATLAS and CMS collaborations, utilising LHC Run 2 data. Here, we investigate the effect of four such ATLAS measurements on parameters sensitive to jet substructure observables. 

\section{Tuning setup}
\label{sec:setup}

The following ATLAS measurements were considered in this study (along with their Rivet identifiers):

\begin{itemize}
    \item Soft-Drop Jet Mass~\cite{Aaboud:2017qwh}(ATLAS\_2017\_I1637587) 
    \item Jet substructure measurements in multijet events~\cite{Aaboud:2019aii} (ATLAS\_2019\_I1724098)
    \item Soft-drop observables~\cite{Aad:2019vyi}(ATLAS\_2019\_I1772062)
    \item Lund jet plane with charged particles~\cite{Aad:2020zcn} (ATLAS\_2020\_I1790256)
    
\end{itemize}

The following parameters were considered in this tuning exercise, with the ranges stated in Table~\ref{tab:my_label6}.

\begin{table}[ht]
        \centering
        \begin{tabular}{lcc}
        \hline
        Parameter  & Lower value & Upper value \\
        \hline
        BeamRemnants:primordialKThard & 1.25 & 3 \\
        ColorReconncetion:range & 1.25 & 3 \\
        TimeShower:pTmin & 0.5 & 1.5  \\
        MultipartonInteractions:pT0Ref & 1.5 & 3 \\
        TimeShower:alphaSvalue & 0.118 & 0.145 \\
        \hline
        \end{tabular}
        \caption{Sampling range of the parameters considered}
        \label{tab:my_label6} 
    \end{table}
    
Weighted \textit{hardQCD} events were generated with a \textit{PThatMin} of 300 GeV. 100 Sampling runs were performed, each with 100000 events. Rivet3~\cite{Bierlich:2019rhm} and Professor tuning system~\cite{Buckley:2009bj} were used. The goodness of sampling and the weight files used can be found in  Appenix~\ref{sec:env} and in Appenix~\ref{sec:wt}.

\FloatBarrier

\section{Results}
\label{sec:res}

The first step was to ascertain where we have a scope of improvement. While a detailed observable-by-observable determination was performed (see  Appendix~\ref{sec:perf}), here we highlight the most salient features:

\begin{itemize}

    \item For Lund Jet Plane (LJP) distributions, we observed that the hard-wide angle emissions part is better modelled by
    the Monash tune whereas the region ranging from UE/MPI to Soft-collinear and Collinear limits are in general better modelled by the A14 tune. However, this distributions also offer the biggest scope of improved modelling.
    \item For the soft drop $\rho$ and $r_g$ observables, in general Monash tune performs somewhat better than A14. 
    One deviation from this trend is when the jet construction is Cluster based, in which case the A14 tune performs better over a large range. 
    \item Both the Jet Substructure and Soft drop jet mass distributions are somewhat better modelled by the A14 tune.
    
\end{itemize}

\begin{figure}[ht]
     \centering
         \includegraphics[width=0.47\textwidth]{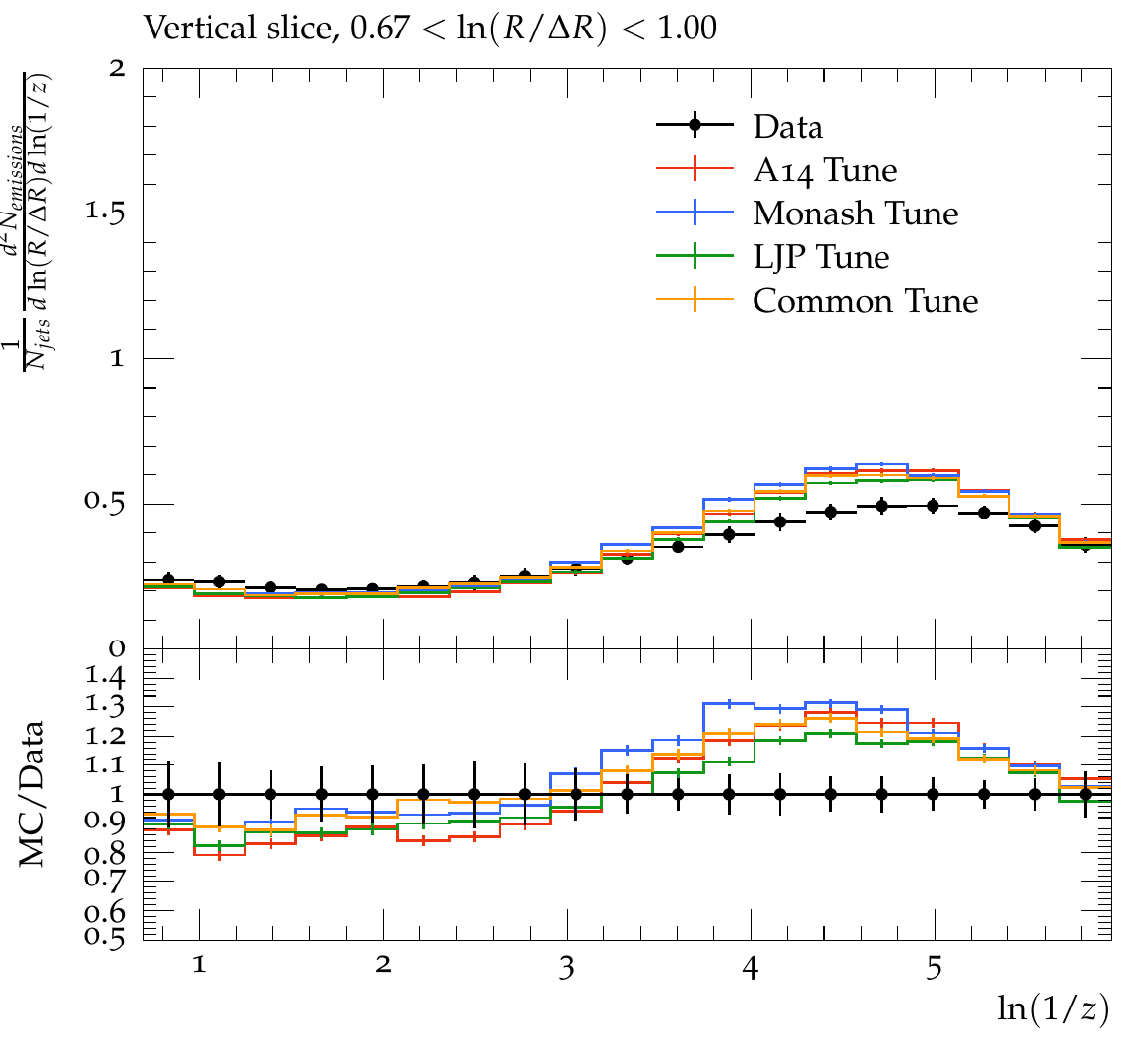}
         \includegraphics[width=0.47\textwidth]{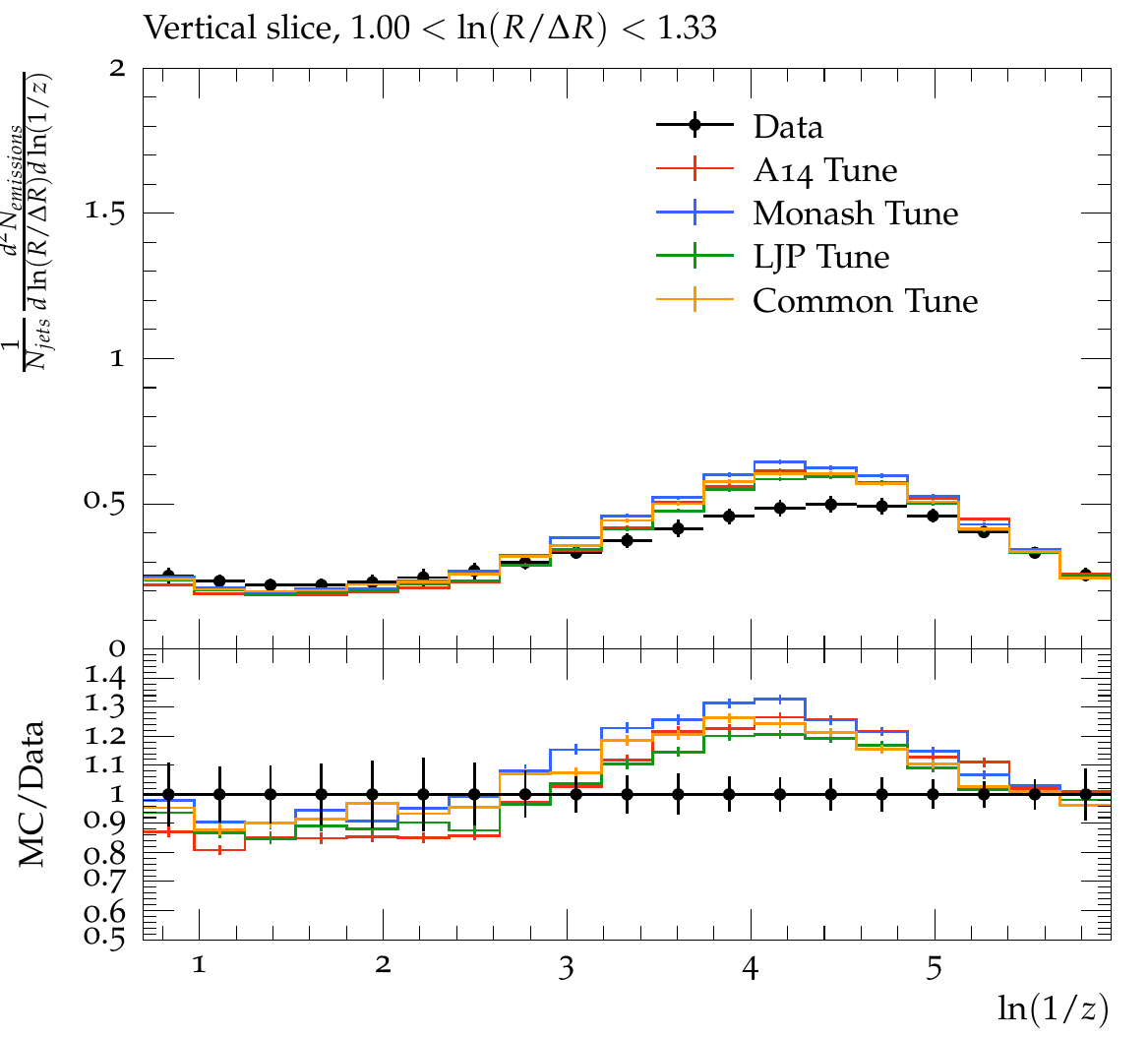}
         \includegraphics[width=0.47\textwidth]{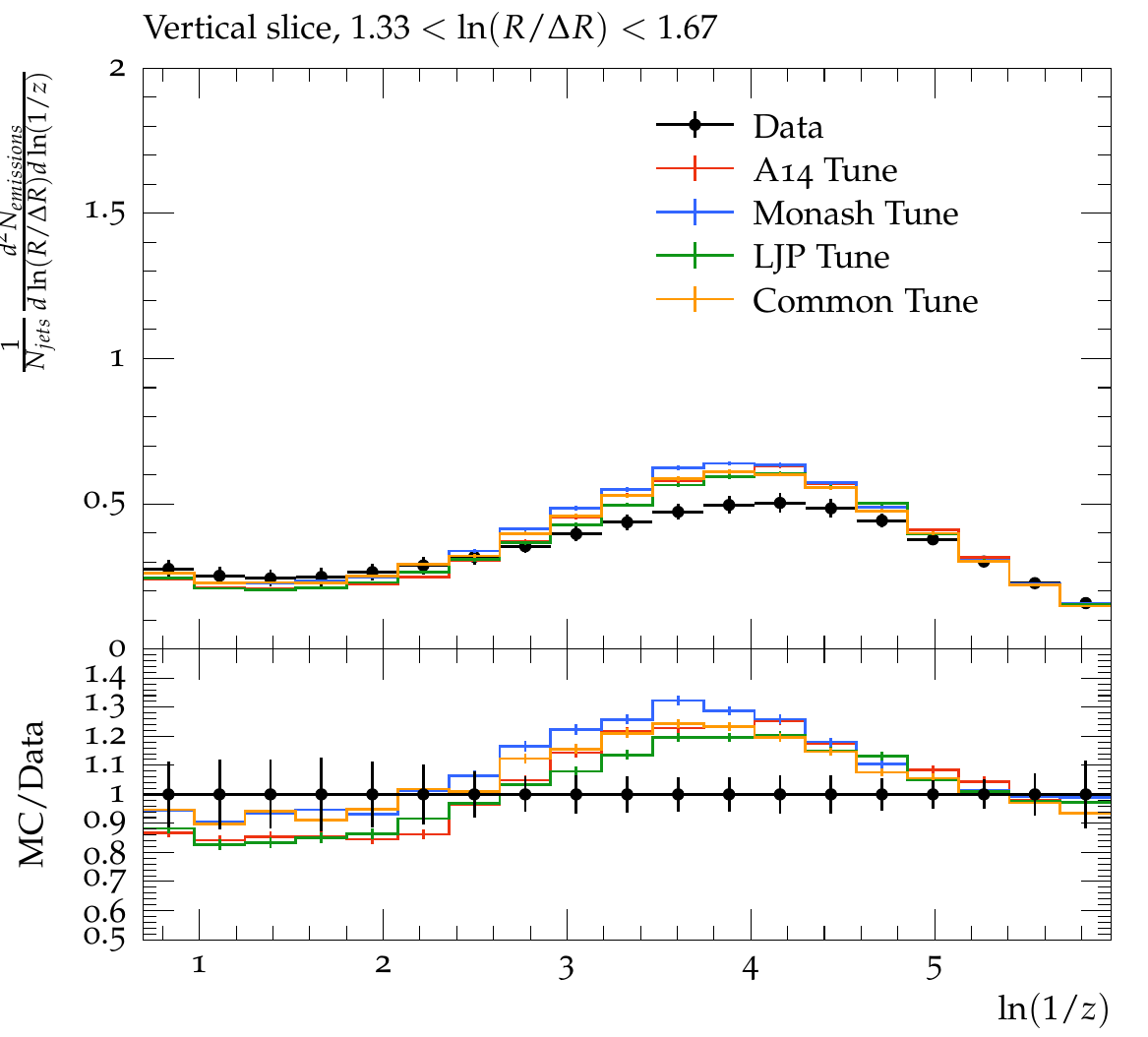}
         \includegraphics[width=0.47\textwidth]{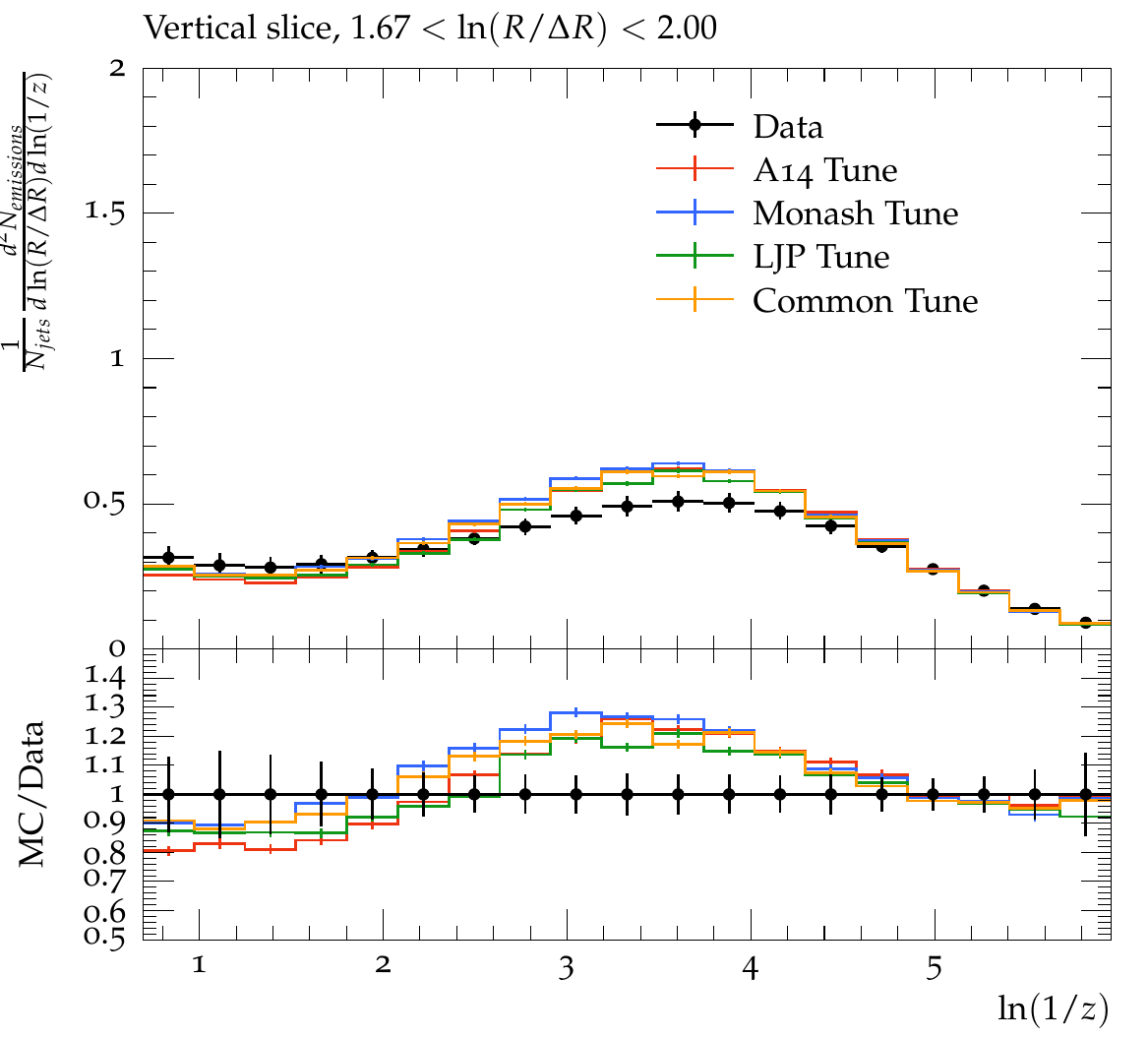}
    \caption{Comparison of our tunes with A14 and Monash tunes for Lund Jet Plane distributions (vertical slices)}
    \label{fig:LJP1}
    \end{figure}

\begin{figure}[ht]
     \centering
         \includegraphics[width=0.47\textwidth]{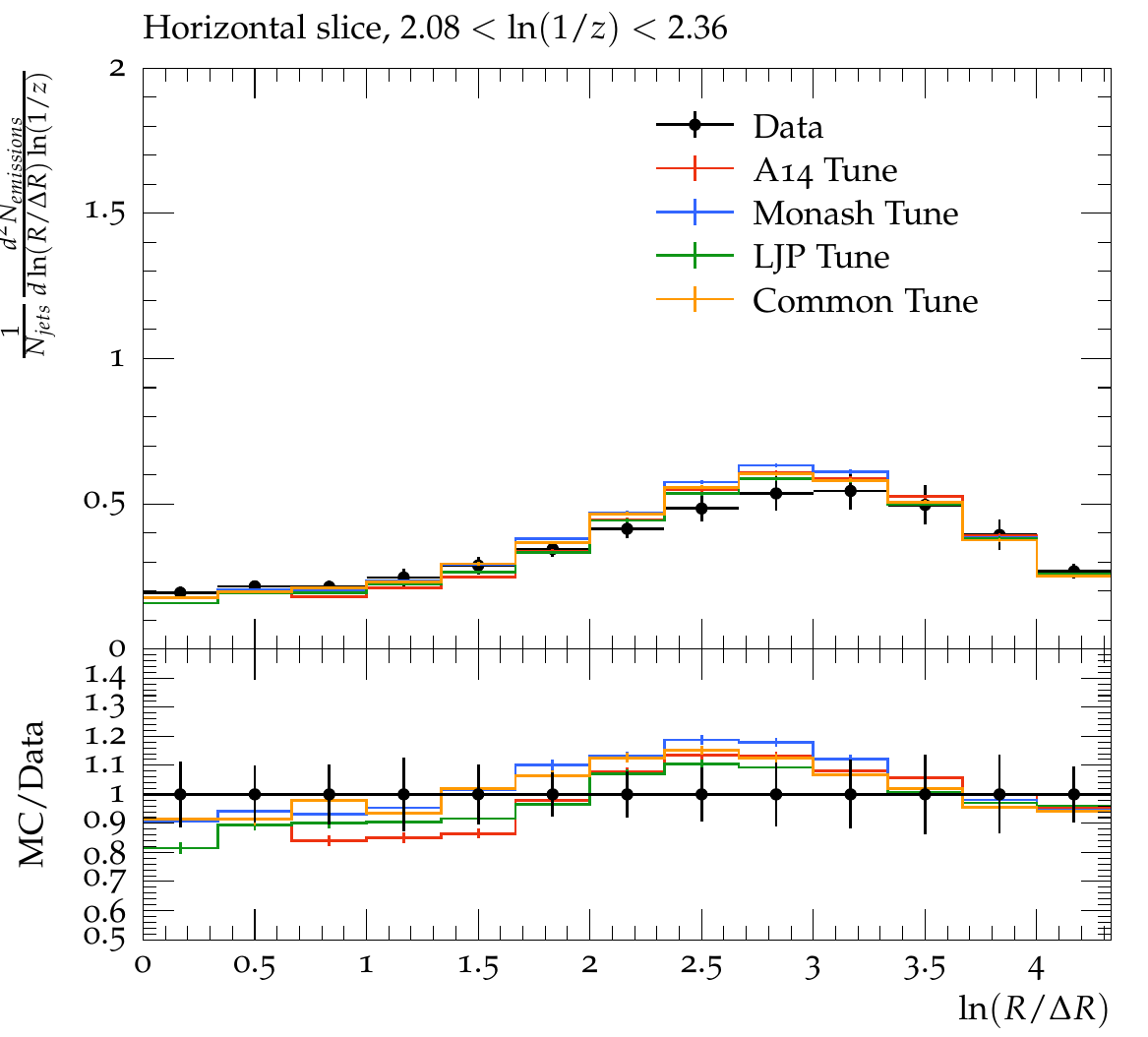}
         \includegraphics[width=0.47\textwidth]{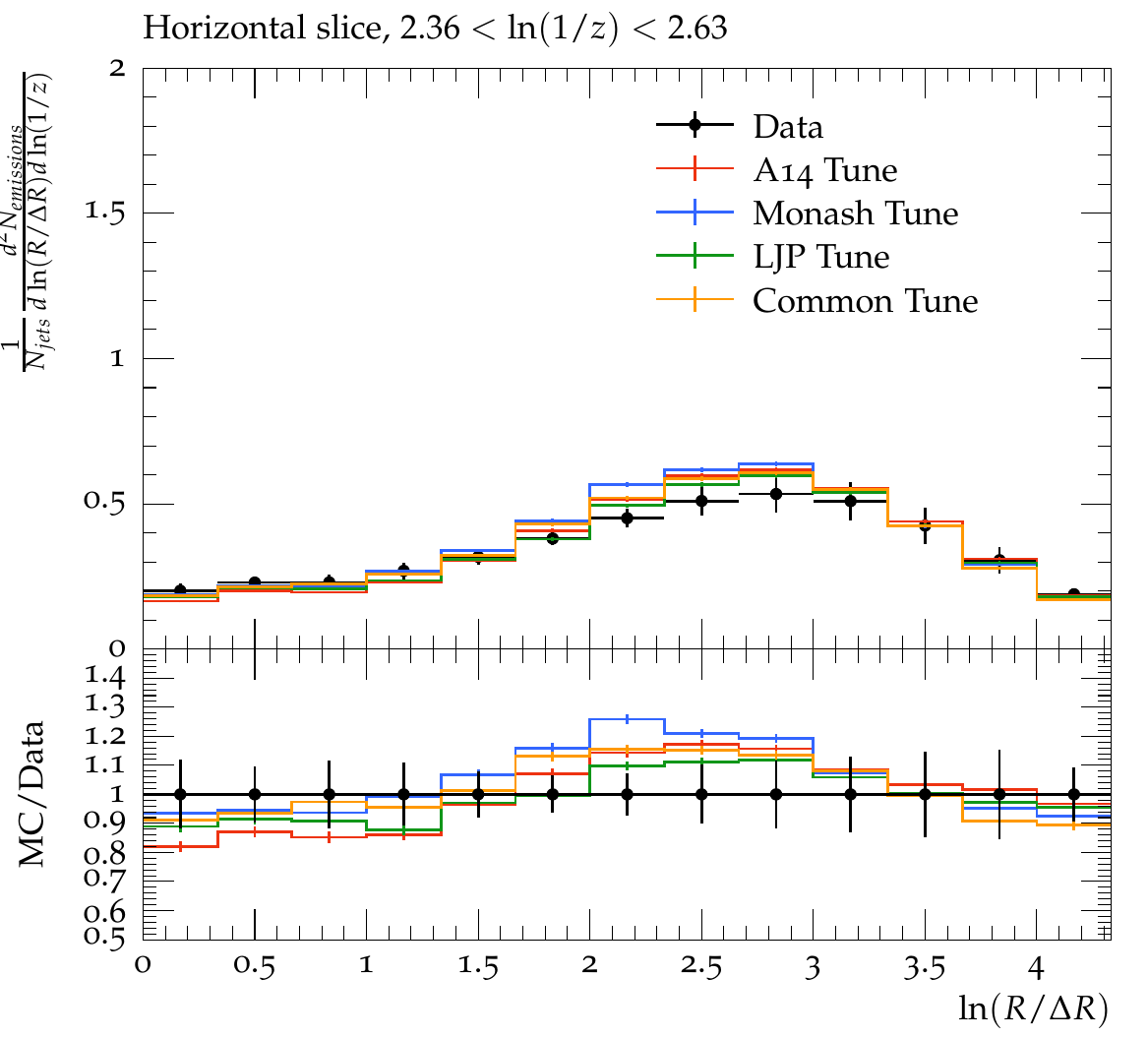}
         \includegraphics[width=0.47\textwidth]{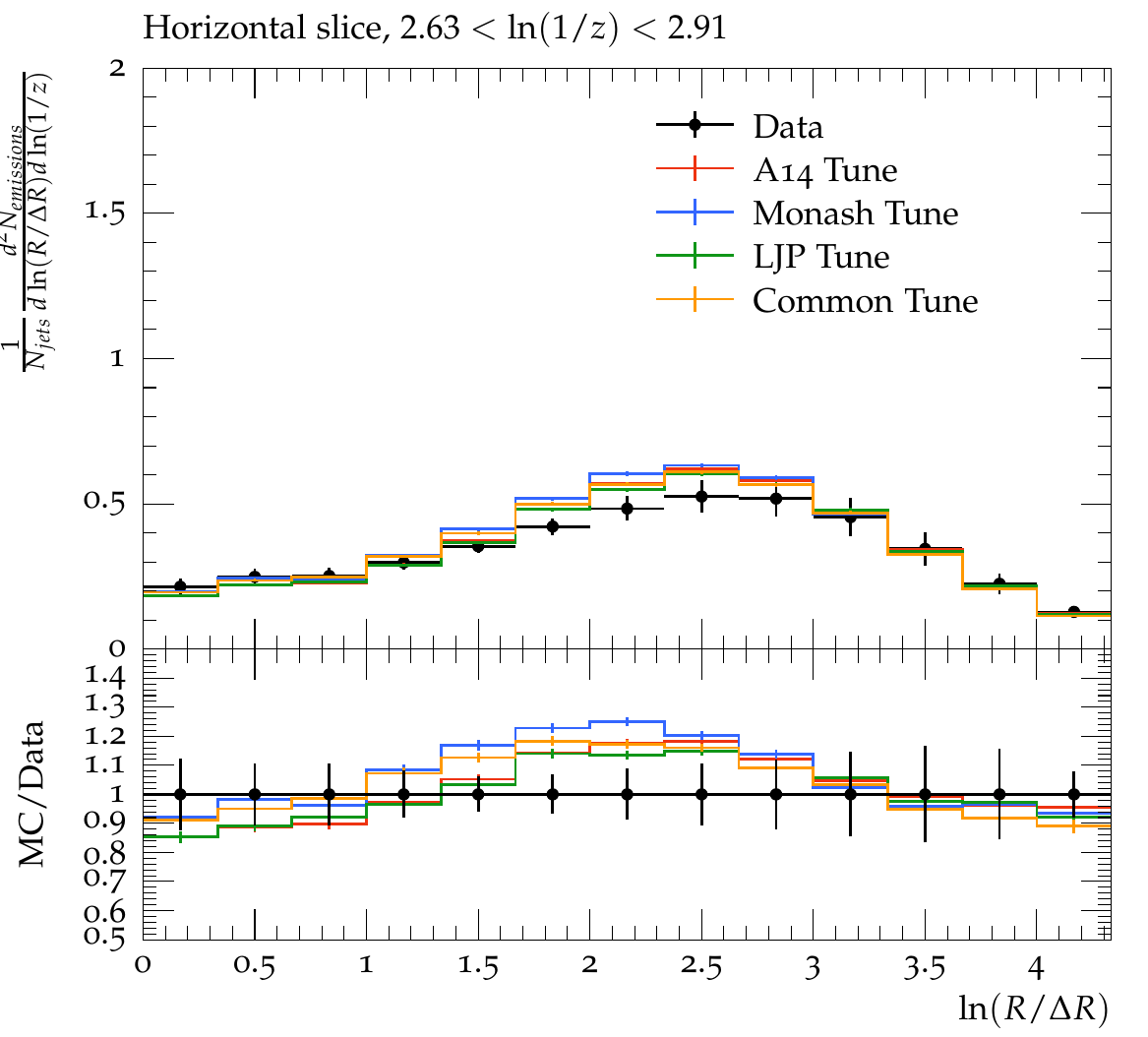}
         \includegraphics[width=0.47\textwidth]{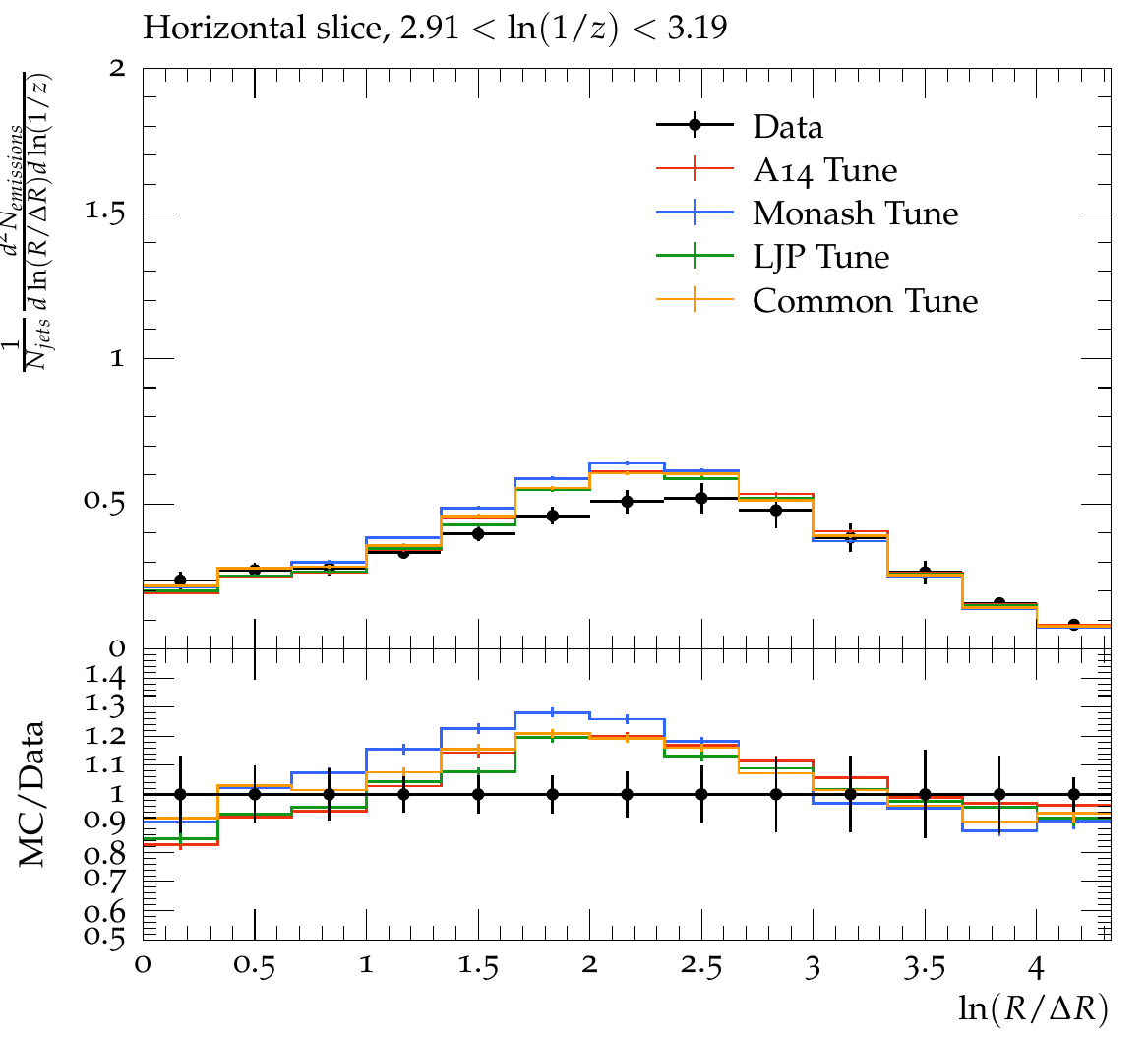}
    \caption{Comparison of our tunes with A14 and Monash tunes for Lund Jet Plane distributions (horizontal slices)}
    \label{fig:LJP2}
    \end{figure}

\begin{figure}[ht]
         \includegraphics[width=0.47\textwidth]{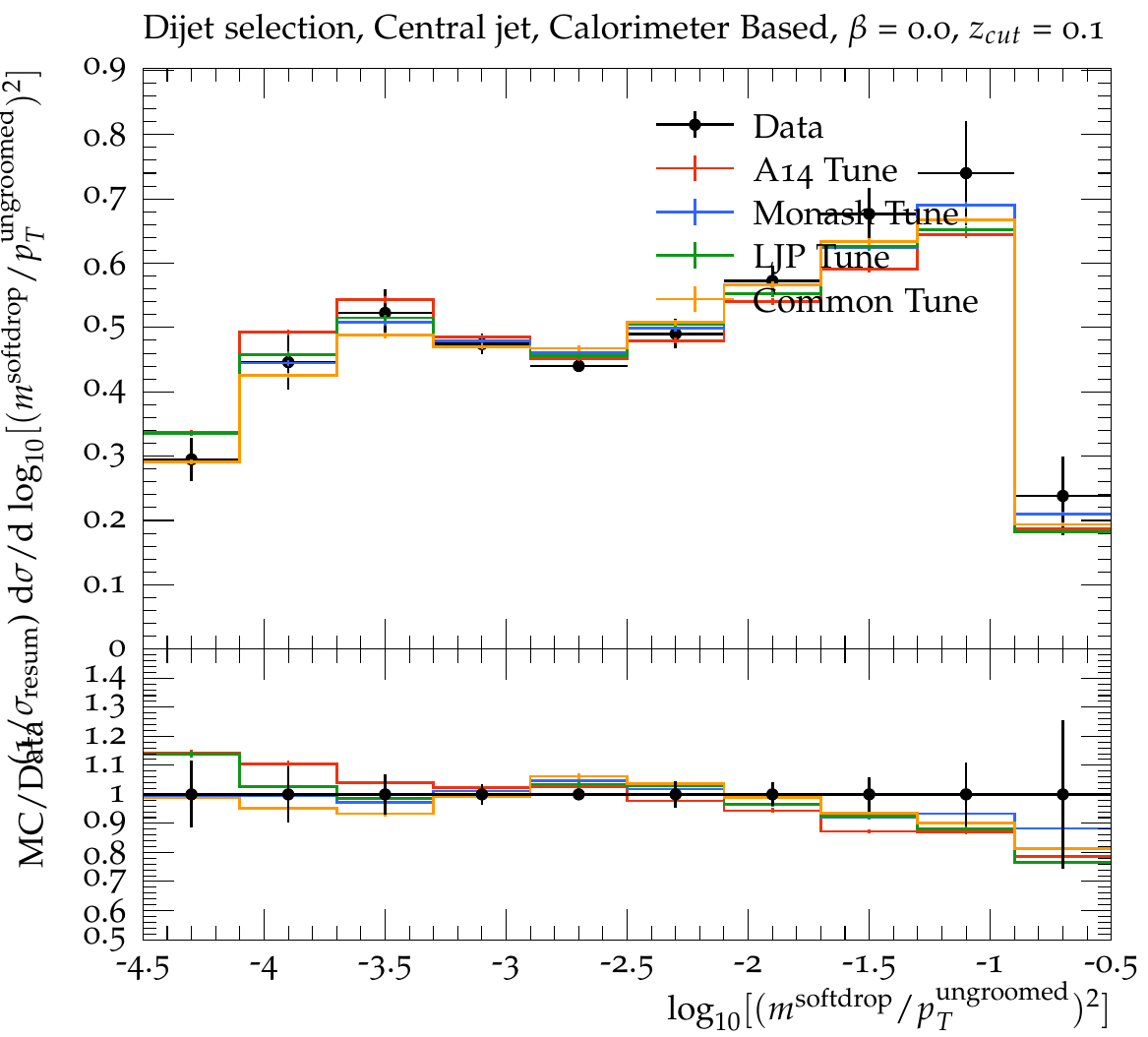}
         \includegraphics[width=0.47\textwidth]{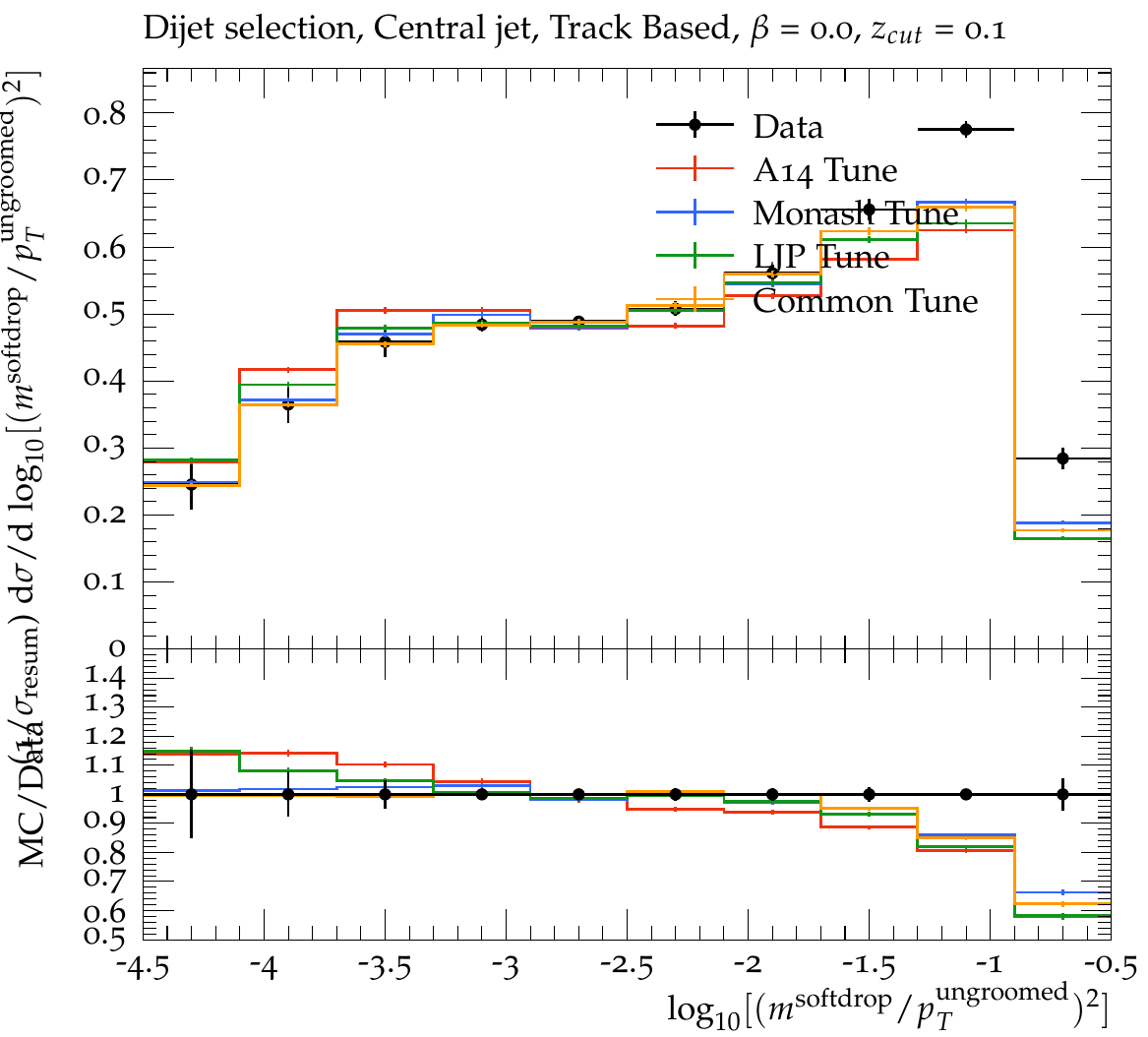}
         \includegraphics[width=0.47\textwidth]{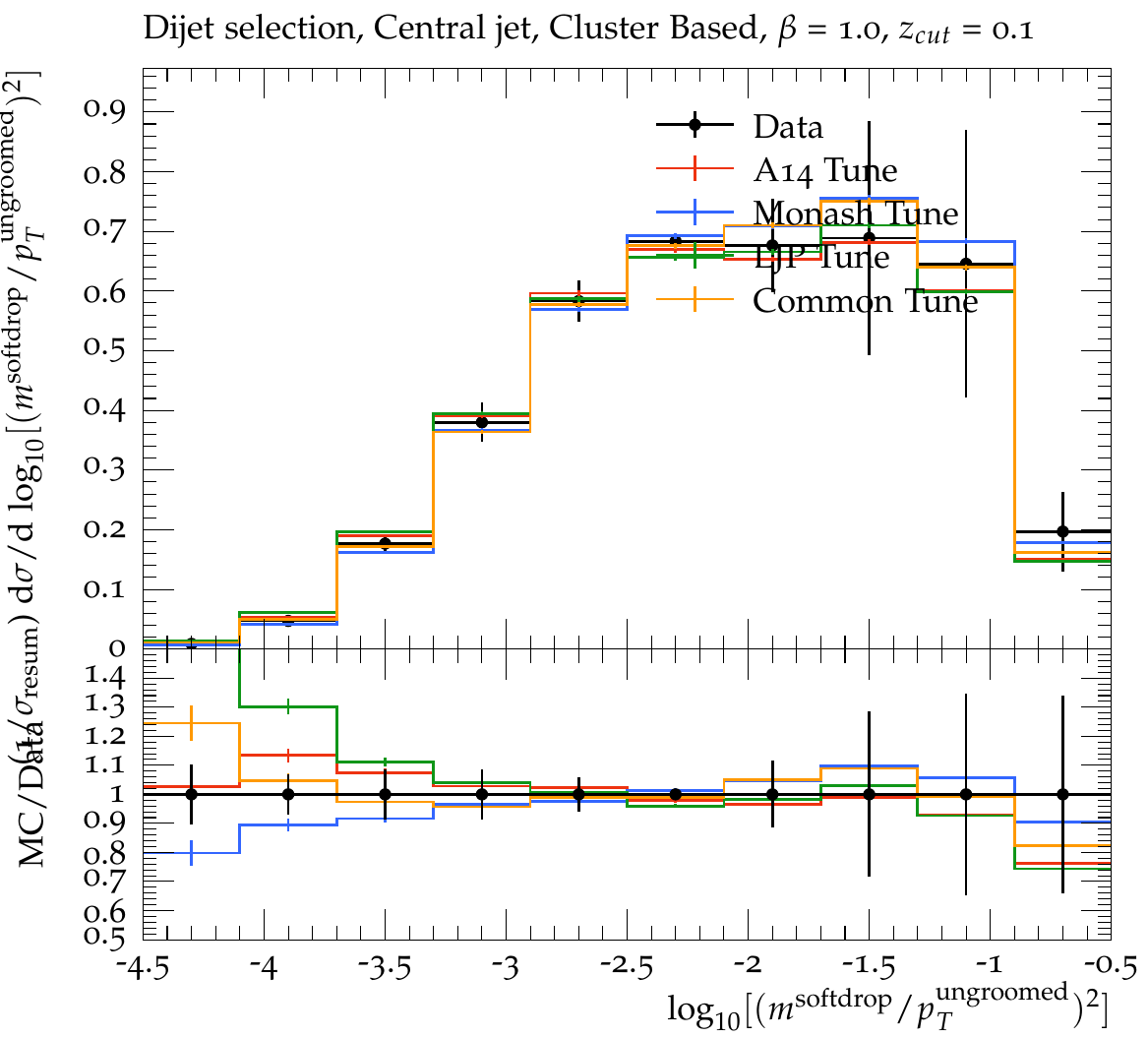}
         \includegraphics[width=0.47\textwidth]{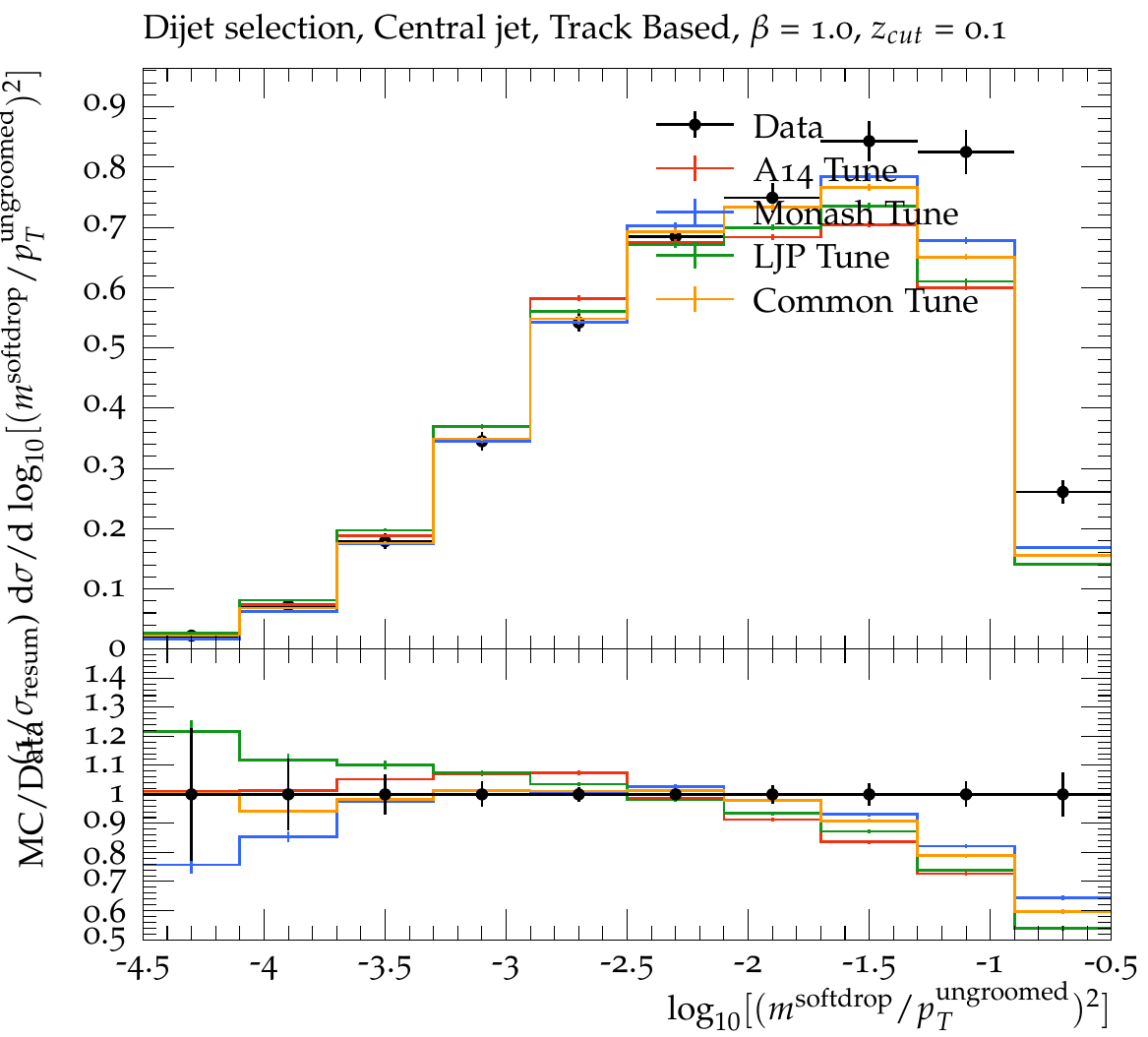}
         \includegraphics[width=0.47\textwidth]{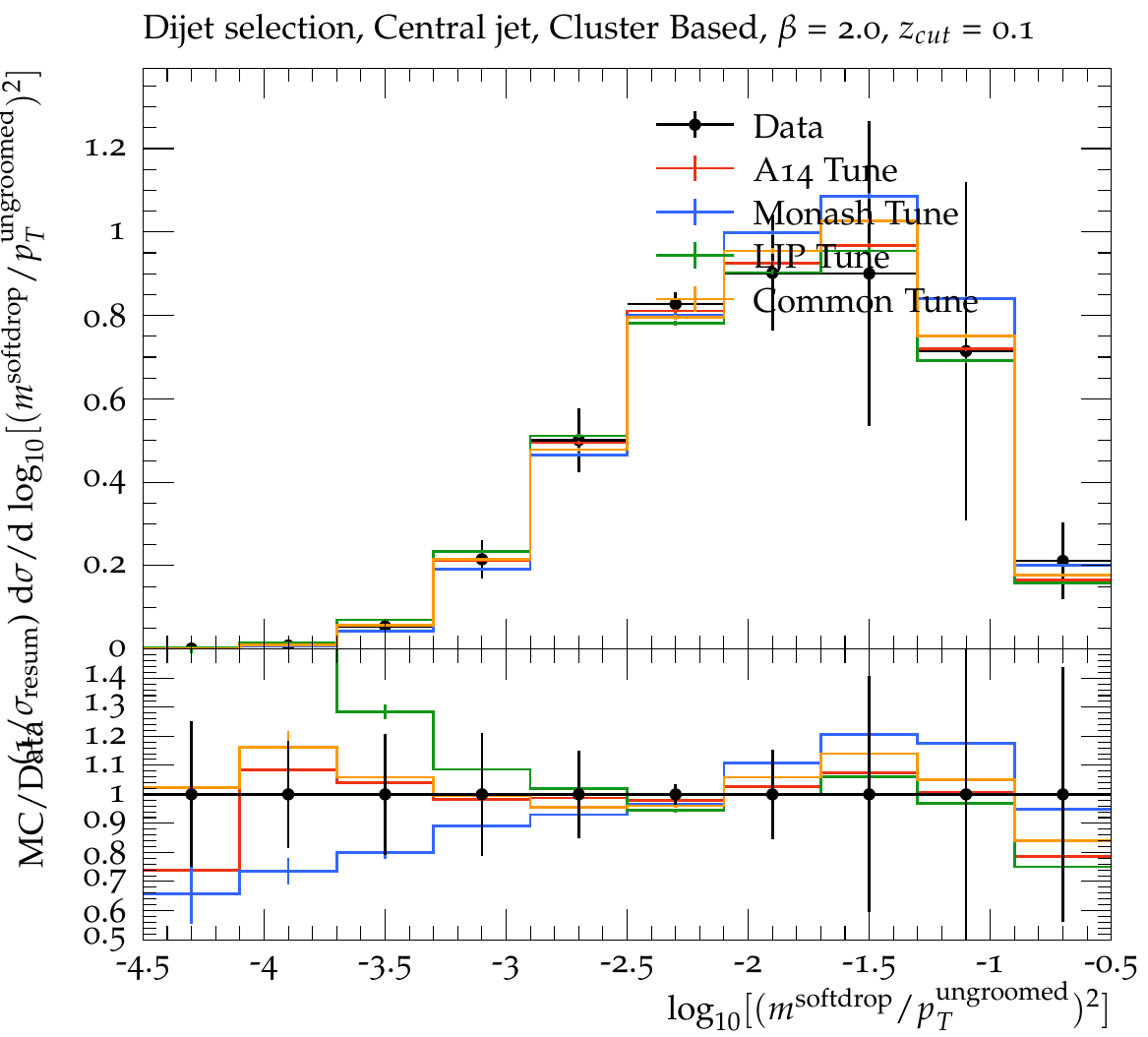}
         \includegraphics[width=0.47\textwidth]{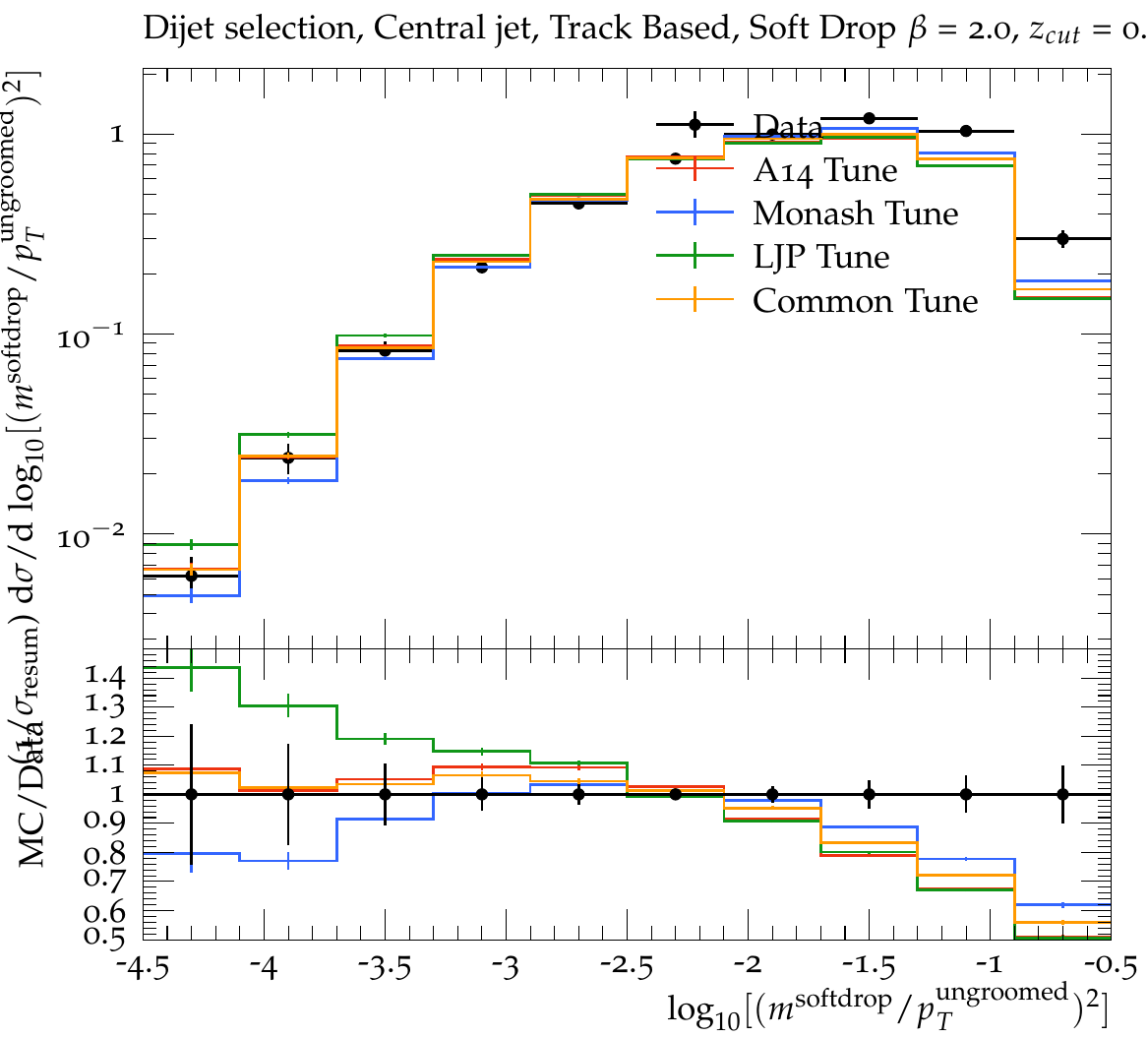}
    \caption{Comparison of our tunes with A14 and Monash tunes for soft drop jet mass distributions}
    \label{fig:SD}
    \end{figure}

\begin{figure}[ht]
         \includegraphics[width=0.47\textwidth]{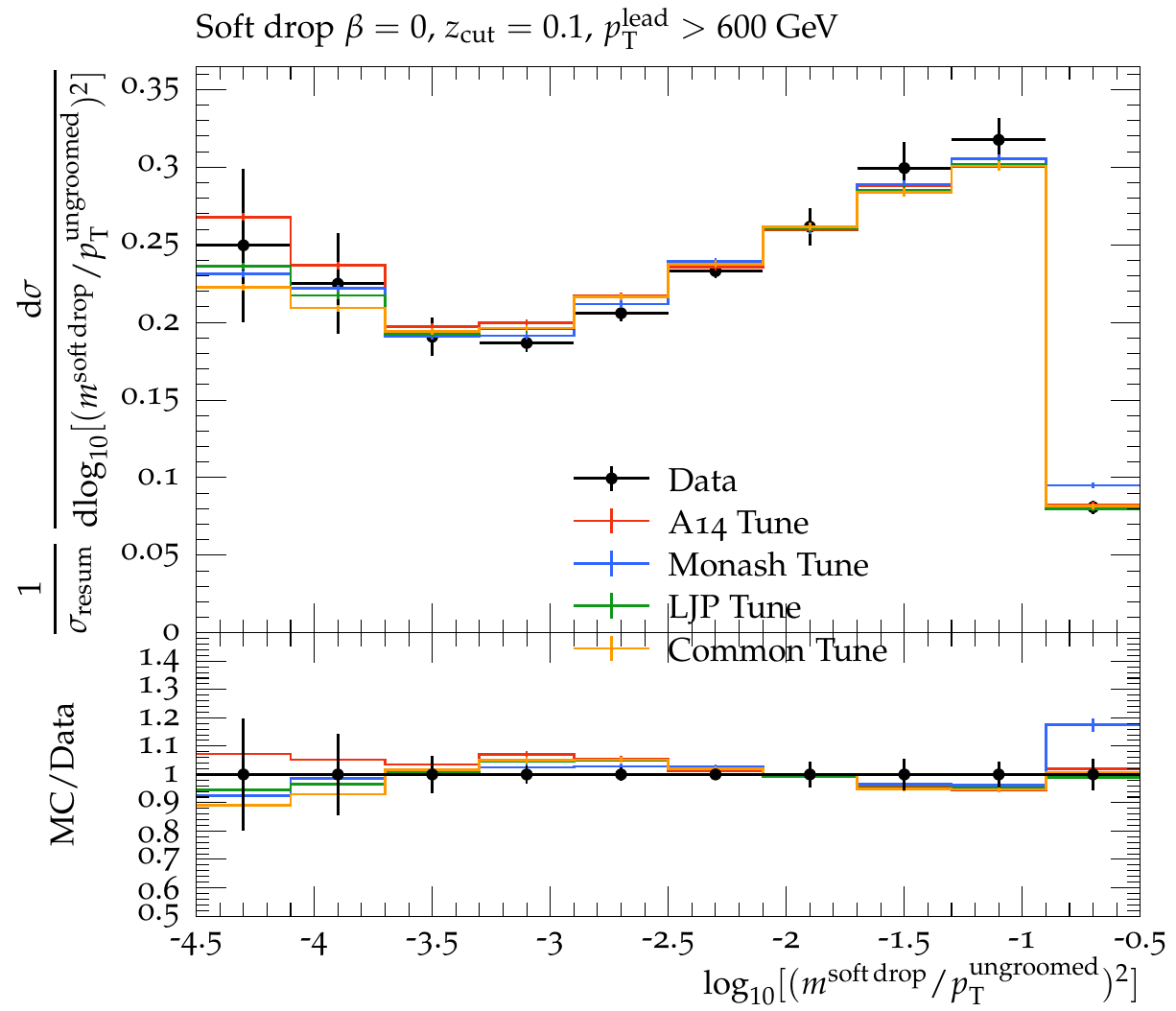}
         \includegraphics[width=0.47\textwidth]{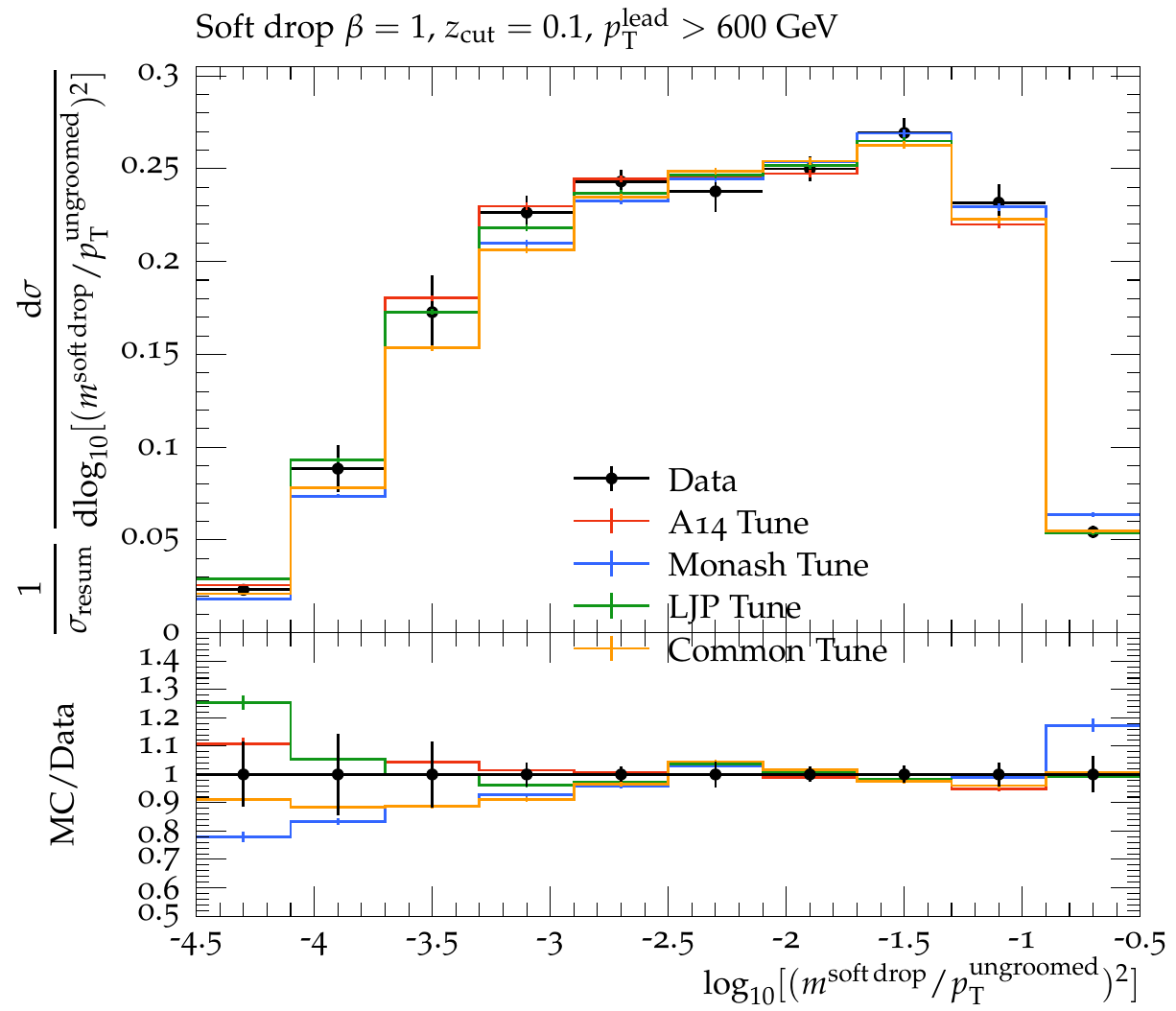}
         \centering
         \includegraphics[width=0.47\textwidth]{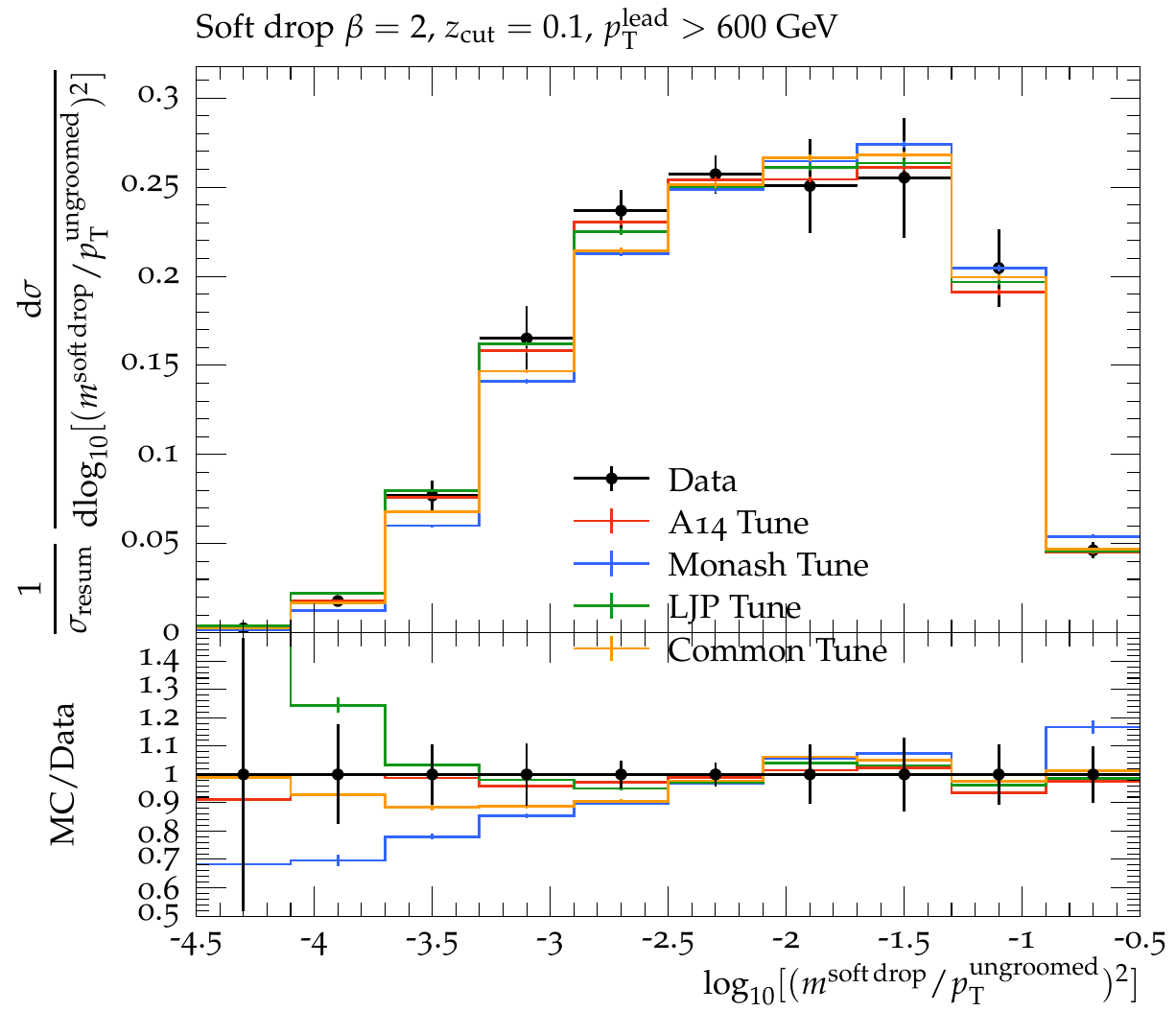}
     \caption{Comparison of our tunes with A14 and Monash tunes for soft drop jet mass distributions}
    \label{fig:sdm}
    \end{figure}

\begin{figure}[ht]
         \includegraphics[width=0.47\textwidth]{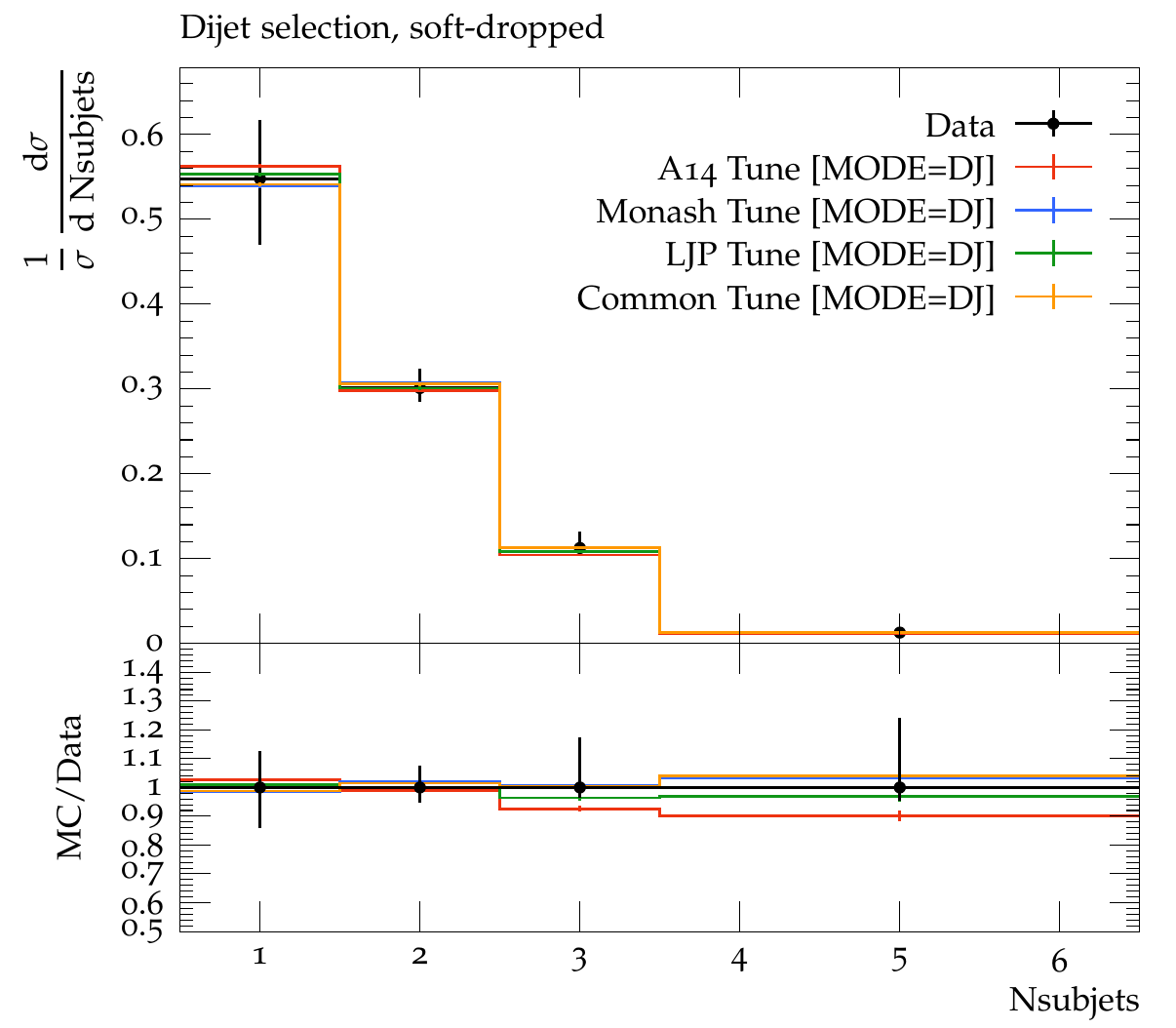}
         \includegraphics[width=0.47\textwidth]{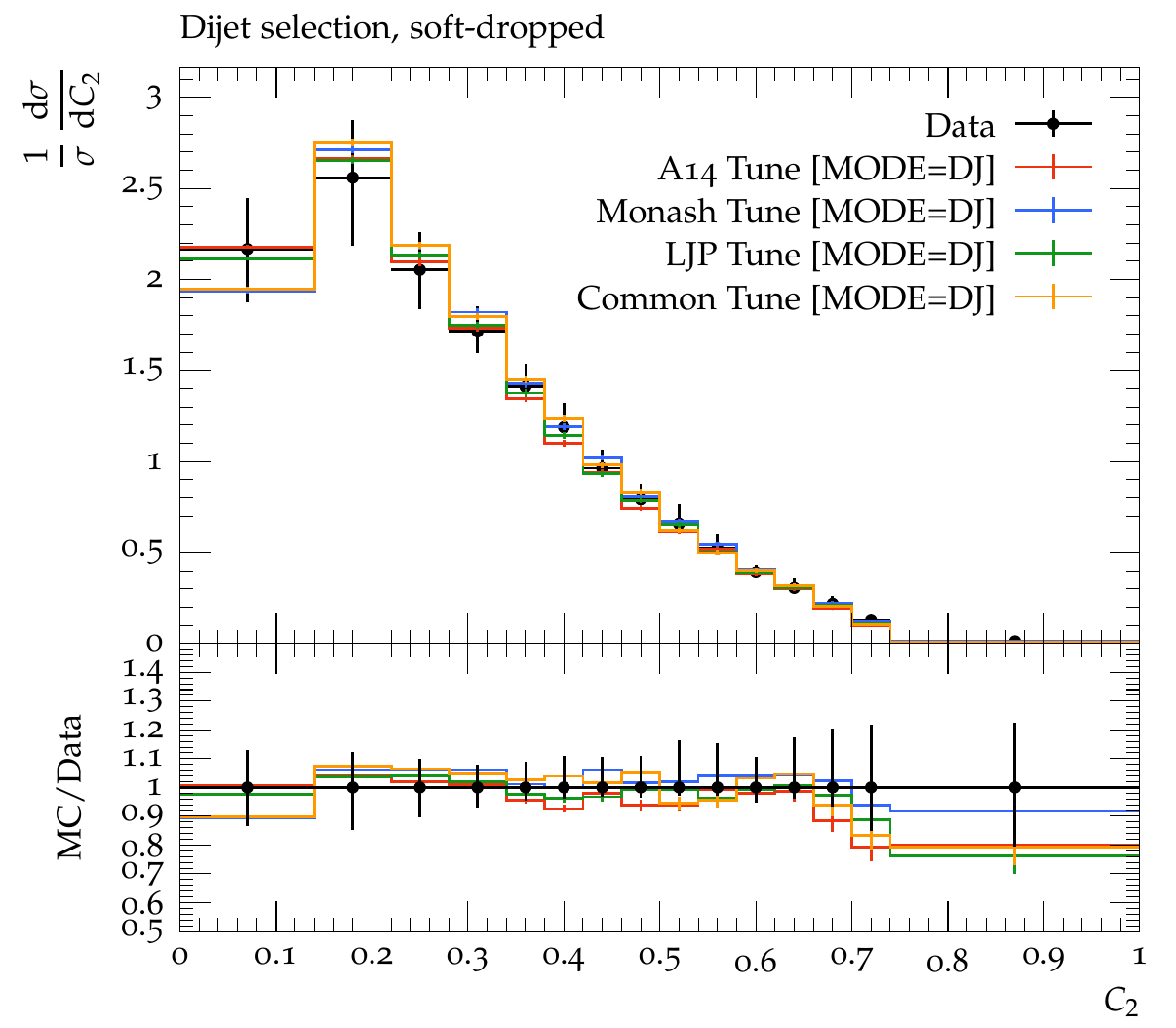}
         \includegraphics[width=0.47\textwidth]{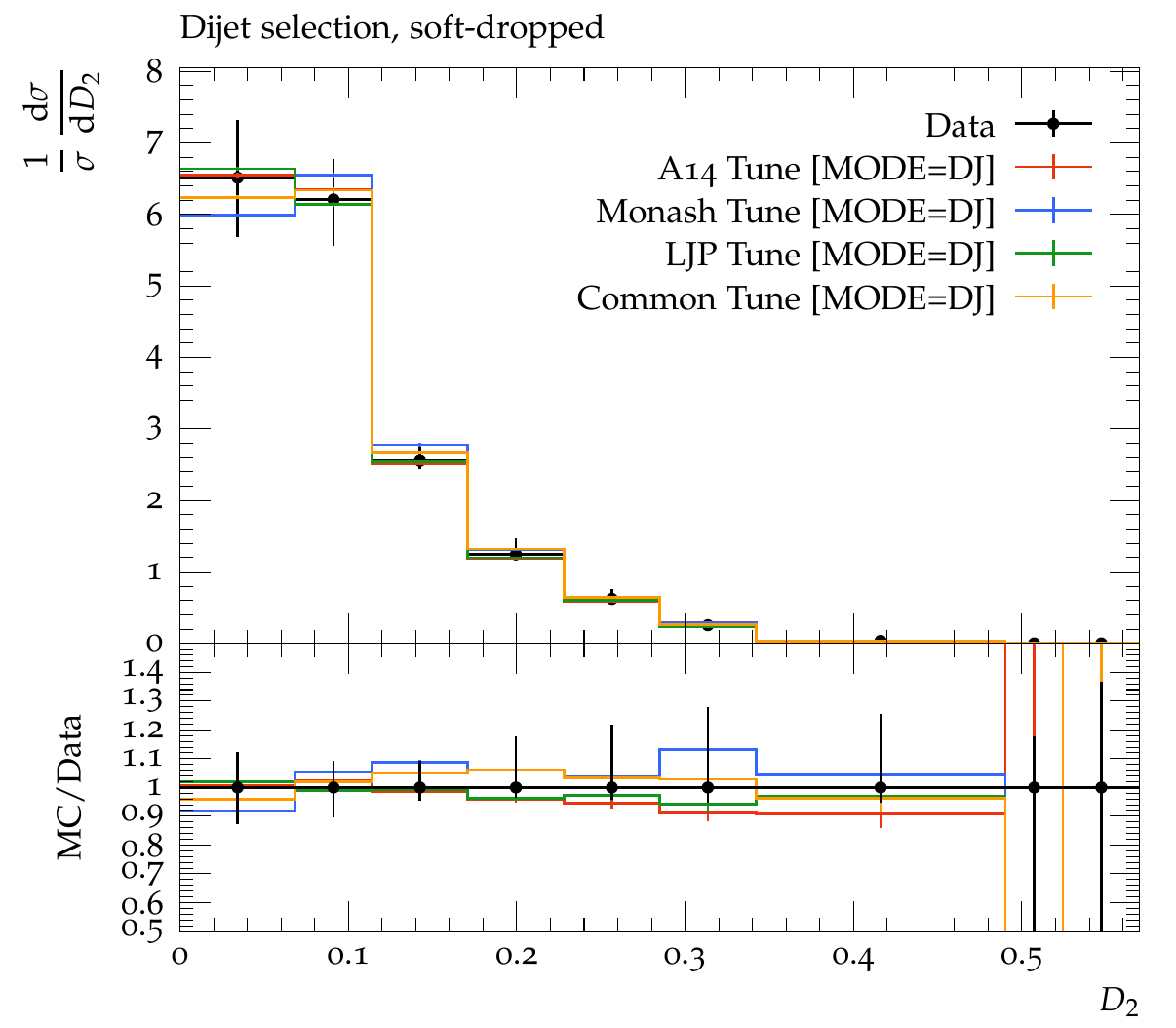}
         \includegraphics[width=0.47\textwidth]{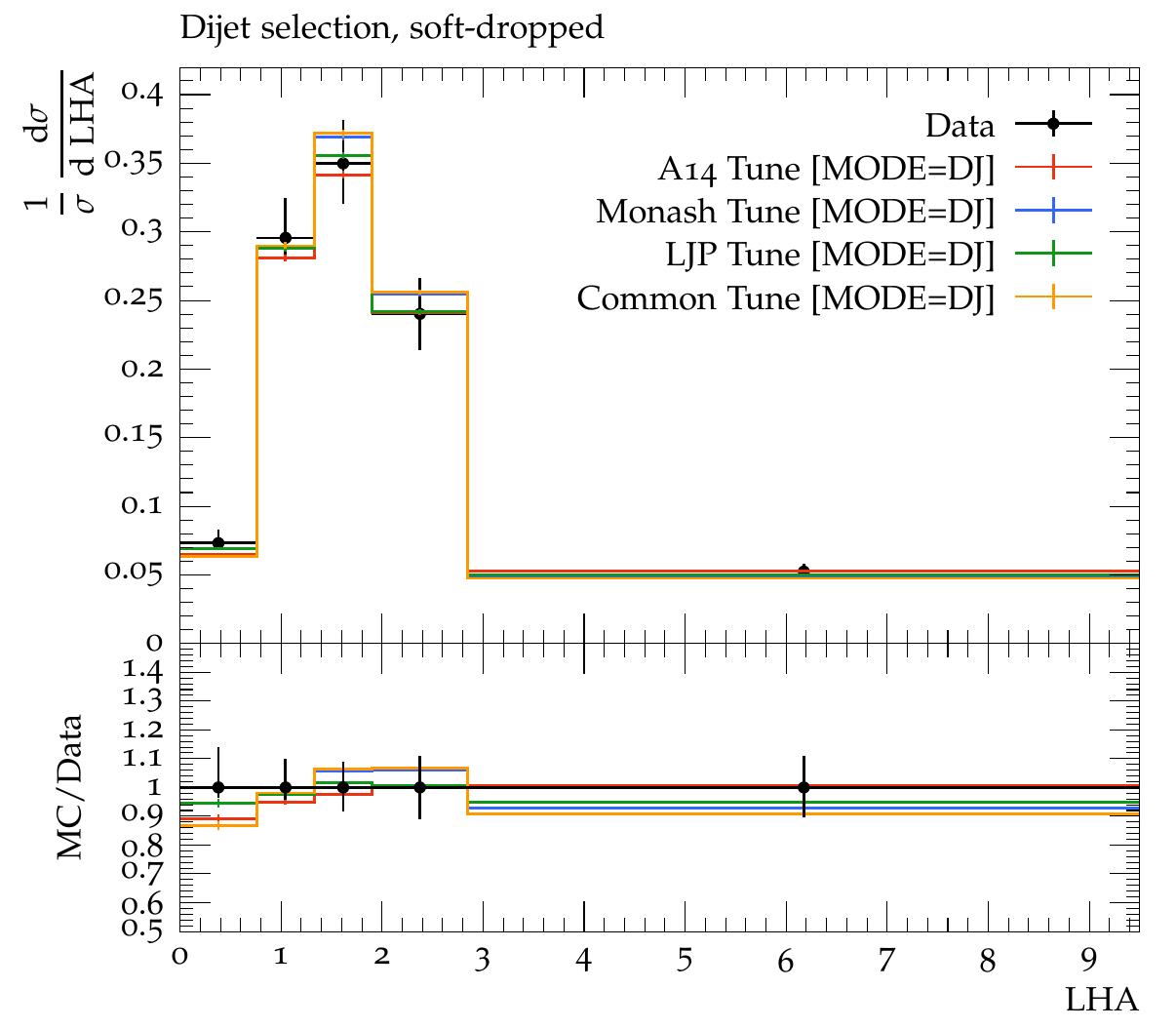}
         \includegraphics[width=0.47\textwidth]{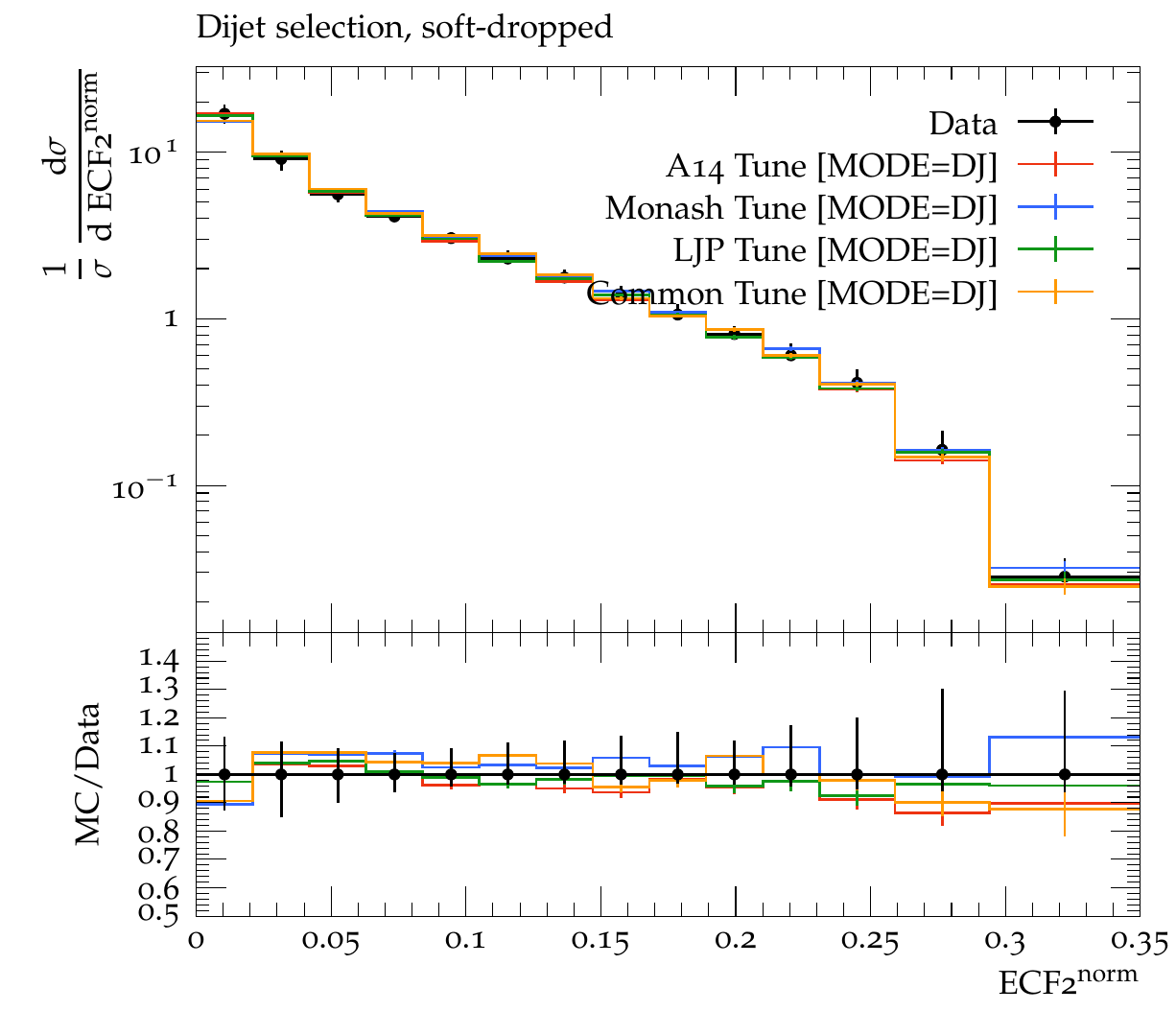}
         \includegraphics[width=0.47\textwidth]{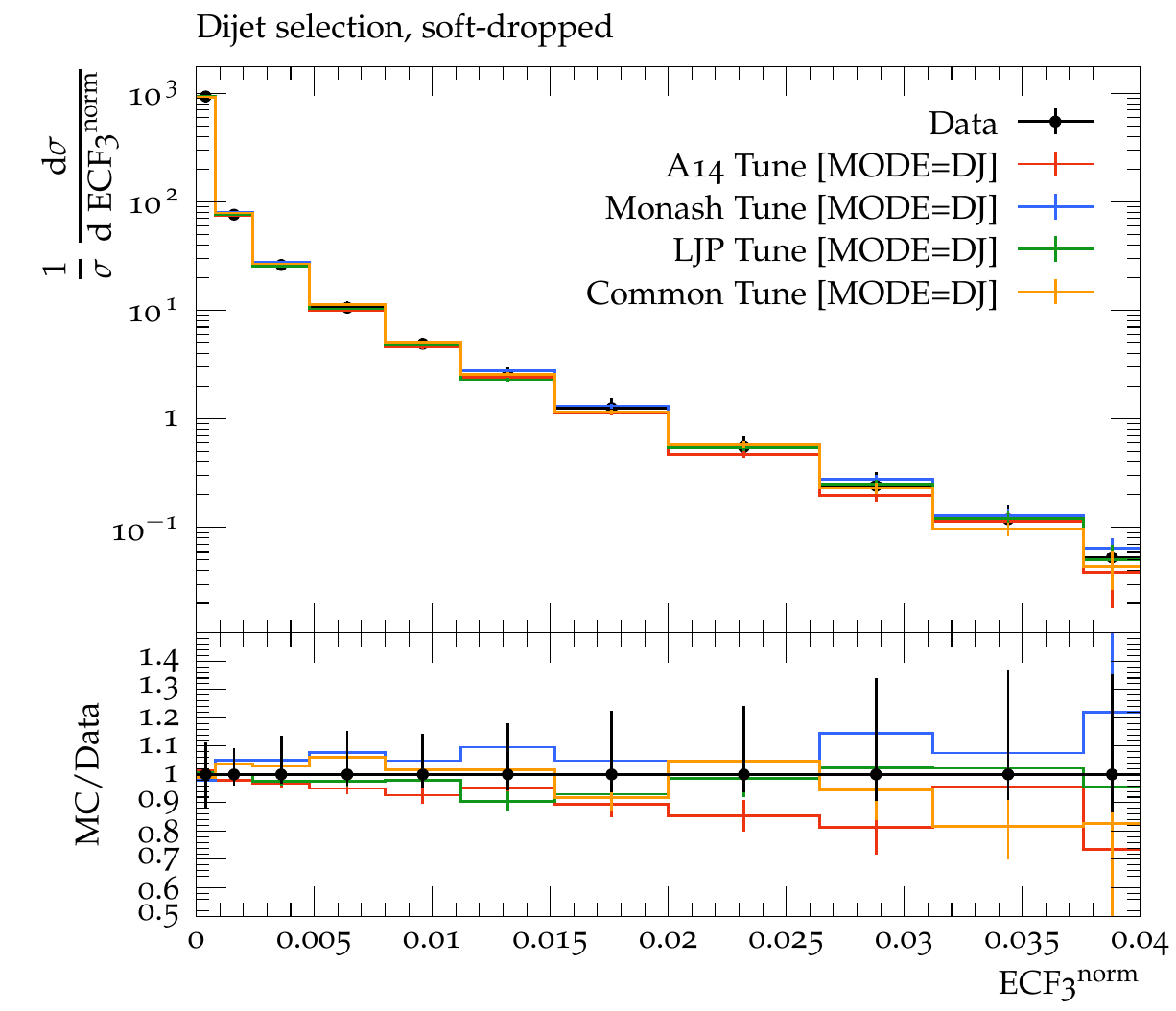}
    \caption{Comparison of our tunes with A14 and Monash tunes for jet Sub structure observable distribution for dijet selection}
    \label{fig:JSS}
    \end{figure}

 Table \ref{tab:result1} lists the parameter values of A14 and Monash along with our tuned values. A separate tune for LJP was performed as this analysis had the largest discrepancy. The LJP tune column shows the parameter values corresponding to the best tune for LJP and the Common Tune column shows the values of the best tune for all the analyses considered. Figures~\ref{fig:LJP1} and \ref{fig:LJP2} show the tuned distributions for the one dimensional vertical slices of the LJP. Figure \ref{fig:SD} shows the tuned distributions for the soft drop observables. Figure \ref{fig:sdm} shows the tuned distributions for soft drop mass. And lastly, Figure \ref{fig:JSS} shows the tuned distributions for the jet substructure observables.

\begin{table}[ht]
        \centering
        \begin{tabular}{lcccc}
        \hline
        Parameters & A14 & Monash & LJP Tune & Common Tune\\
        \hline
        BeamRemnants:primordialKThard & 1.88 & 1.8 & 2.288 & 2.065\\
        ColorReconnection:range & 1.71 & 1.8 & 2.73 & 1.69\\
        TimeShower:pTmin & 0.40 & 0.50 & 1.288 & 0.775\\
        MultipartonInteractions:pT0Ref & 2.09 & 2.28 & 2.766&  2.91\\
        TimeShower:alphaSvalue & 0.127 & 0.1365 & 0.1308 & 0.1309\\
        \hline
        \end{tabular}
        \caption{Comparison of tuned values with Monash and A14}
        \label{tab:result1}
\end{table}

\FloatBarrier

\section{Summary}
\label{sec:summ}
   
The results obtained show small improvements of roughly 5-10\% in the distributions of the Lund Jet Plane and Soft Drop Mass distributions from the previous A14 and Monash Tunes. As in Table \ref{tab:result1}, it can be seen that the parameter values of the tunes obtained are pulled up from the A14 and Monash Tunes. In the case of the LJP , we see that the A14 and Monash Tunes deviate most from the data near the peaks of the distributions. This is the region where soft collinear effects transitions to UE/MPI effects in the LJP. Since the tunes we obtained improve this region of the distributions, it can be inferred that higher values of these parameters facilitate more soft radiations in the final state.

In the case of the soft drop observable distributions, there are regions that require generation of more mass to model the data better. These compete with the LJP values and decreases values of parameters: \textit{BeamRemnants:primordialKThard} from 2.288 to 2.065, \textit{ColorReconnection:range} from 2.73 to 1.69, \textit{TimeShower:pTmin} from 1.288 to 0.775. For the other two parameters, \textit{MPI:pT0Ref} and \textit{TimeShower:alphaSvalue}, the values increased slightly.

\section*{Acknowledgements}

DK is funded by National Research  Foundation (NRF), South Africa through Competitive Programme for Rated Researchers (CPRR), Grant No: 118515. We thank Andy Buckley and Holger Schulz for technical assistance with Professor program, as well as for physics discussions.

\bibliographystyle{elsarticle-num.bst}
\bibliography{ref.bib}

\newpage

\section{Appendix}

\subsection{Tune performance}
\label{sec:perf}

\begin{table}[ht]
\scriptsize
        \centering
        \begin{tabular}{cccllc}
        \hline
        Plots & Observable & $\ln(R/\Delta R)$ or & \multicolumn{2}{c}{Better Tune Performance regions} & Better Tune\\
        \cline{4-5}
          & & $\ln(1/z)$ Slice & \textbf{A14} & \textbf{Monash} & (overall)\\
        \hline
        d03-x01-y01 & $\ln(1/z)$ & 0.00-0.33 & 4.3-6 & 0.7-4.3 & Monash\\
        \hline
        d04-x01-y01 & $\ln(1/z)$ & 0.33-0.67 & 3.4-4.2 & 0.7-3.4 , 4.4-6 & Monash\\
        \hline
        d05-x01-y01 & $\ln(1/z)$ & 0.67-1.00 & 3-5 & 0.7-3 , 5-6 & Monash\\
        \hline
        d06-x01-y01 & ln(1/z) & 1.00-1.33 & 3-4.2 & 0.7-3 , 4.2-6 & Monash\\
        \hline
        d07-x01-y01 & $\ln(1/z)$ & 1.33-1.67 & 2.4-4.2 & 0.7-2.4 , 4.3-6 & Monash \\
        \hline
        d08-x01-y01 & $\ln(1/z)$ & 1.67-2.00 & 2-4 , 5.2 , 5.8 & 0.7-2 , 4-5 , 5.5 & -\\
        \hline
        d09-x01-y01 & $\ln(1/z)$ & 2.00-2.33 & 1.8-4 , 4.7, 5.2-6 & 0.7-1.8, 4-4.4 & A14\\
        \hline
        d10-x01-y01 & $\ln(1/z)$ & 2.33-2.67 & 1.6-3, 4.4-5.8 & 0.7-1.4, 3.2-4.2 & A14\\
        \hline
        d11-x01-y01 & $\ln(1/z)$ & 2.67-3.00 & 1.4-5 & 0.7-1.2, 5.2-5.8 & A14\\
        \hline
        d12-x01-y01 & $\ln(1/z)$ & 3.00-3.33 & 0.8-1.4, 3-4, 4.7, 5.8 & 1.6-3, 3.3, 4-4.5, 5-5.6 & -\\
        \hline
        d13-x01-y01 & $\ln(1/z)$ & 3.33-3.67 & 0.7-3.4, 4.6-5 & 3.6-4.4, 5.1-6 & A14\\
        \hline
        d14-x01-y01 & $\ln(1/z)$ & 3.67-4.00 & 0-3.4 & 3.6-4.6 & A14\\
        \hline
        d15-x01-y01 & $\ln(1/z)$ & 4.00-4.33 & 0.7-2.6, 3.6, 4-4.6, 5, 5.57 & 2.7-3.4, 3.9, 4.7, 5.23& A14\\
        \hline
        d16-x01-y01 & $\ln(R/\Delta R)$ & 0.69-0.97 & 3-4.5 &  0-3 & Monash\\
        \hline
        d17-x01-y01 & $\ln(R/\Delta R)$ & 0.97-1.25 & 3-4 & 0-3 & Monash\\
        \hline
        d18-x01-y01 & $\ln(R/\Delta R)$ & 1.25-1.52 & 2.7-3.2, 3.5-4 & 0-2.6 & Monash\\
        \hline
        d19-x01-y01 & $\ln(R/\Delta R)$ & 1.52-1.80 & 2.4-4.5 & 0.5-2.3 & -\\
        \hline
        d20-x01-y01 & $\ln(R/\Delta R)$ & 1.80-2.08 & 2-3, 3.3-4.5 & 0-2 & - \\
        \hline
        d21-x01-y01 & $\ln(R/\Delta R)$ & 2.08-2.36 & 1.6-4.5 & 0-1.6, 2.2, 3.4-4 & -\\
        \hline
        d22-x01-y01 & $\ln(R/\Delta R)$ & 2.36-2.63 & 1.4-3, 3.6-4.5 & 0-1.3, 3-3.6 & A14\\
        \hline
        d23-x01-y01 & $\ln(R/\Delta R)$ & 2.63-2.91 & 1.4-3, 3.5 & 0-1.3, 3-4.5 & -\\
        \hline
        d24-x01-y01 & $\ln(R/\Delta R)$ & 2.91-3.19 & 0.6-4.5 & 0-0.5 & A14\\
        \hline
        d25-x01-y01 & $\ln(R/\Delta R)$ & 3.19-3.47 & 0.6-2.4, 3.4-4.5 & 0-0.5, 2.4-3.3 & A14\\
        \hline
        d26-x01-y01 & $\ln(R/\Delta R)$ & 3.47-3.74 & 0.5-4.5 & 0.2, 1.2, 2.5, 3.5 & A14\\
        \hline
        d27-x01-y01 & $\ln(R/\Delta R)$ & 3.74-4.02 & 0.3-4.5 & 0.2 & A14\\
        \hline
        d28-x01-y01 & $\ln(R/\Delta R)$ & 4.02-4.30 & 0-1.6, 3.7-4.5 & 1.7-3.6 & -\\
        \hline
        d29-x01-y01 & $\ln(R/\Delta R)$ & 4.30-4.57 & 0.2, 0.5,1.2, 2.3-4.5,  & 0.8, 1-2.2, 3.2, 3.8  & -\\
        \hline
        d30-x01-y01 & $\ln(R/\Delta R)$ & 4.57-4.85 & 0.2, 1.6-3.3, 3.5 & 0.3-1.5, 3.7-4.5 & -\\
        \hline
        d31-x01-y01 & $\ln(R/\Delta R)$ & 4.85-5.13 & 0-3, 3.4-4.5 & 3.2 & A14\\
        \hline
        d32-x01-y01 & $\ln(R/\Delta R)$ & 5.13-5.41 & 0.2, 1.7-4 & 0.4-1.7, 2.75 & A14\\
        \hline
        d33-x01-y01 & $\ln(R/\Delta R)$ & 5.41-5.68 & 0.2, 2-3 & 0.5-2, 3.2-4 & Monash\\
        \hline
        d34-x01-y01 & $\ln(R/\Delta R)$ & 5.68-5.96 & 1.7-2.3 & 0-1.7, 2.5-3 & Monash\\
        \hline
        \end{tabular}
        \caption{ATLAS\_2020\_I1790256(LJP)}
        \label{tab:t1}
    \end{table}

 \begin{table}[ht]
        \scriptsize
        \centering
        \begin{tabular}{cccccllc}
        \hline
        Plots  &  & $\beta$ & $z_{cut}$ & Observable & \textbf{A14} & \textbf{Monash} & Better Tune \\
        \hline
        d01-x01-y01 & Calorimeter based &0 & 0.1 & $\rho$ & - & all & Monash\\
        \hline
        d02-x01-y01 & Track based & 0 & 0.1 & $\rho$ & - &  all & Monash\\
        \hline
        d03-x01-y01 & Cluster based & 1 & 0.1 & $\rho$ &  [-4.5,-3.7], & [-3.5,-2.1], & A14/\\
        &&&&& [-2,-1.3] & [-1,-0.5] & Monash \\
        \hline 
         d04-x01-y01 & Track based & 1 & 0.1 & $\rho$ &  [-4.5,-3.7] & [-3.5,-0.5] & Monash\\
        \hline
        d05-x01-y01 & Cluster based & 2 & 0.1 & $\rho$ &  [-4.5,-1.1] & -0.7 & A14\\
        \hline
        d06-x01-y01 & Track based & 2 & 0.1 & $\rho$ &  [-4.5,-3.7] & [-3.5,-0.5] & Monash\\
        \hline
        d07-x01-y01 & Track based & 1 & 0.1 & $\rho$ &  [-4.5,-3.7] & [-3.5,-0.5] & Monash\\
        \hline
        d16-x01-y01 & Track based & 1 & 0.1 & $r_{g}$ &  - & all & Monash\\
        \hline
        d17-x01-y01 & Cluster based & 2 & 0.1 & $r_{g}$ &  [-1.2,-0.2] & -0.15 & A14\\
        \hline
        d18-x01-y01 & Track based & 2 & 0.1 & $r_{g}$ &  -1.1 & [-1,-0.1] & Monash\\
        \hline
        d19-x01-y01 & Central jet/Calorimeter & 0 & 0.1 & $r_{g}$ &  - & all  & Monash\\
        \hline
        d20-x01-y01 & Central jet/Track & 0 & 0.1 & $r_{g}$ & - & all & Monash\\
        \hline
        d21-x01-y01 & Central jet/Cluster & 1 & 0.1 & $\rho$  & [-4.5,-1] & -0.7 & A14\\
        \hline
        d22-x01-y01 & Central jet/Track & 1 & 0.1 & $\rho$  & [-4.5,-3.7]& [-3.5,-0.5] & Monash\\
        \hline
        d23-x01-y01 & Central jet/Cluster & 2 & 0.1 & $\rho$  & [-3.5,-0.9]& -0.7 & A14\\
        \hline
        d24-x01-y01 & Central jet/Track & 2 & 0.1 & $\rho$  & [-4.5,-3.7]& [-3.5,-0.7] & Monash\\
        \hline
        d34-x01-y01 & Central jet/Track & 1 & 0.1 & $r_{g}$ & - & all & Monash\\
        \hline
        d35-x01-y01 & Central jet/Cluster & 2 & 0.1 & $r_{g}$ & [-1.2,-0.4] & -0.5,-0.15 & A14\\
        \hline
        d36-x01-y01 & Central jet/Track & 2 & 0.1 & $r_{g}$ & -1.1 & [-1,-0.1] & Monash\\
        \hline
        d37-x01-y01 & Forward jet/Calorimeter & 0 & 0.1 & $r_{g}$ & - & all & Monash\\
        \hline
        d38-x01-y01 & Forward jet/Track & 0 & 0.1 & $\rho$  & - & all & Monash\\
        \hline
        d39-x01-y01 & Forward jet/Cluster & 1 & 0.1 & $\rho$  & -4.3 & [-4,-0.5] & Monash\\
        \hline
        d40-x01-y01 & Forward jet/Track & 1 & 0.1 & $\rho$ & [-4.5,-3.7] & [-3.5,-0.5] & Monash\\
        \hline
        d41-x01-y01 & Forward jet/Cluster & 2 & 0.1 & $\rho$ & -3.9,[-3.1,-1] & -3.5,-0.7 & A14\\
        \hline
        d42-x01-y01 & Forward jet/Track & 2 & 0.1 & $\rho$ & [-4.5,-3.7] & [-3.5,-0.5] & Monash\\
        \hline
        d49-x01-y01 & Forward jet/Track & 0 & 0.1 & $r_{g}$ & all & all & -\\
        \hline
        d51-x01-y01 & Forward jet/Cluster & 1 & 0.1 & $r_{g}$ & [-0.8,-0.2] & [-1.2,-0.8],-0.1 & -\\
        \hline
        d52-x01-y01 & Forward jet/Track & 1 & 0.1 & $r_{g}$ & - & all & Monash\\
        \hline
        d53-x01-y01 & Forward jet/Cluster & 2 & 0.1 & $r_{g}$ & all & -0.15 & A14\\
        \hline
        d54-x01-y01 & Forward jet/Track & 2 & 0.1 & $r_{g}$ & -1.1 & [-1,-0.1] & Monash\\
        \hline
        \end{tabular}
        \caption{ATLAS 2019 I1772062(Soft\_Drop\_Jet Observables)}
        \label{tab:t2}
    \end{table}

  \begin{table}[ht]
    \footnotesize
        \centering
        \begin{tabular}{ccllc}
        \hline
        Plots & Observable & \multicolumn{2}{c}{Better Tune Performance regions} & Better Tune\\
        \cline{3-4}
          & & \textbf{A14} & \textbf{Monash} & (overall)\\
        \hline
        d01-x01-y01 & Nsubjets & 0-10 & - & A14 \\
        \hline
        d02-x01-y01 & $C_2$ & 0-0.86 & 0.36-0.42 , 0.64-0.72 & A14\\
        \hline
        d03-x01-y01 & $D_2$ & 0-0.5 & - & A14\\
        \hline
        d04-x01-y01 & LHA & 0,4.5 & - & A14\\
        \hline
        d05-x01-y01 & ECF$_2^{norm}$ & 0-0.252 & 0.252-0.35 & A14\\
        \hline
        d06-x01-y01 & ECF$_3^{norm}$ & 0-0.04 & - & A14\\
        \hline
        d23-x01-y01 & Nsubjets & 1-2 & 2-5 & Monash \\
        \hline
        d24-x01-y01 & $C_2$ & 0-0.38 , 0.42-0.5 , 0.54-0.62 & 0.4 , 0.52 , 0.62-1.0 & A14\\
        \hline
        d25-x01-y01 & $D_2$ & 0-0.48 & - & A14\\
        \hline
        d26-x01-y01 & LHA & 0-1.4 & 1.4-9 & A14\\
        \hline
        d27-x01-y01 & ECF$_2^{norm}$ & 0-0.23 & ECF$_2^{norm}$ 0.23-0.35 & A14\\
        \hline
        d28-x01-y01 & ECF$_3^{norm}$ & 0-0.02 & 0.02-0.32 & A14\\
        \hline
        \end{tabular}
        \caption{ATLAS 2019 I1724098(JSS, Dijet Selection)}
        \label{tab:t3}
    \end{table}

    \begin{table}[ht]
    \footnotesize
        \centering
        \begin{tabular}{cccllc}
        \hline
        Plots & Observable & $\beta$ & \multicolumn{2}{c}{Better Tune Performance regions} & Better Tune\\
        \cline{4-5}
         & ($p_T^{lead}>600$ GeV) & ($z_{cut} = 0.1$) & \textbf{A14} & \textbf{Monash} & (overall)\\
        \hline
        d01-x01-y01 & $\log_{10}$[($m^{\textrm{soft drop}}/p_{T}^{\textrm{ungroomed}})^2$] & 0 & [-2.5,-0.5] & [-4,-2.5] & A14\\
        \hline
        d02-x01-y01 & $\log_{10}$[($m^{\textrm{soft drop}}/p_{T}^{\textrm{ungroomed}})^2$] & 1 & [-4.5,-0.5] & - & A14\\
        \hline
        d03-x01-y01 & $\log_{10}$[($m^{\textrm{soft drop}}/p_{T}^{\textrm{ungroomed}})^2$] & 2 & [-4.5,-0.5] & - & A14\\
        \hline
        \end{tabular}
        \caption{ATLAS 2017 I1637587(Soft Drop Mass)}
        \label{tab:t4}
    \end{table}

\FloatBarrier

\subsection{Envelope plots}
\label{sec:env}

Having decided the range of values for each parameter, we visualise the region of the distributions to check its utility and hence it is important before proceeding further. This can be done using the PROFESSOR tool by generating envelopes plots with the comand {\fontfamily{qcr}\selectfont prof2-envelopes}. The envelope plots show an area covering the distributions which indicates the bin values that the observables can take within the selected parameter ranges. These are shown in the following sub sections.

\subsubsection{Soft Drop Mass Distributions}

    \begin{figure}[ht]
         \includegraphics[width=0.47\textwidth]{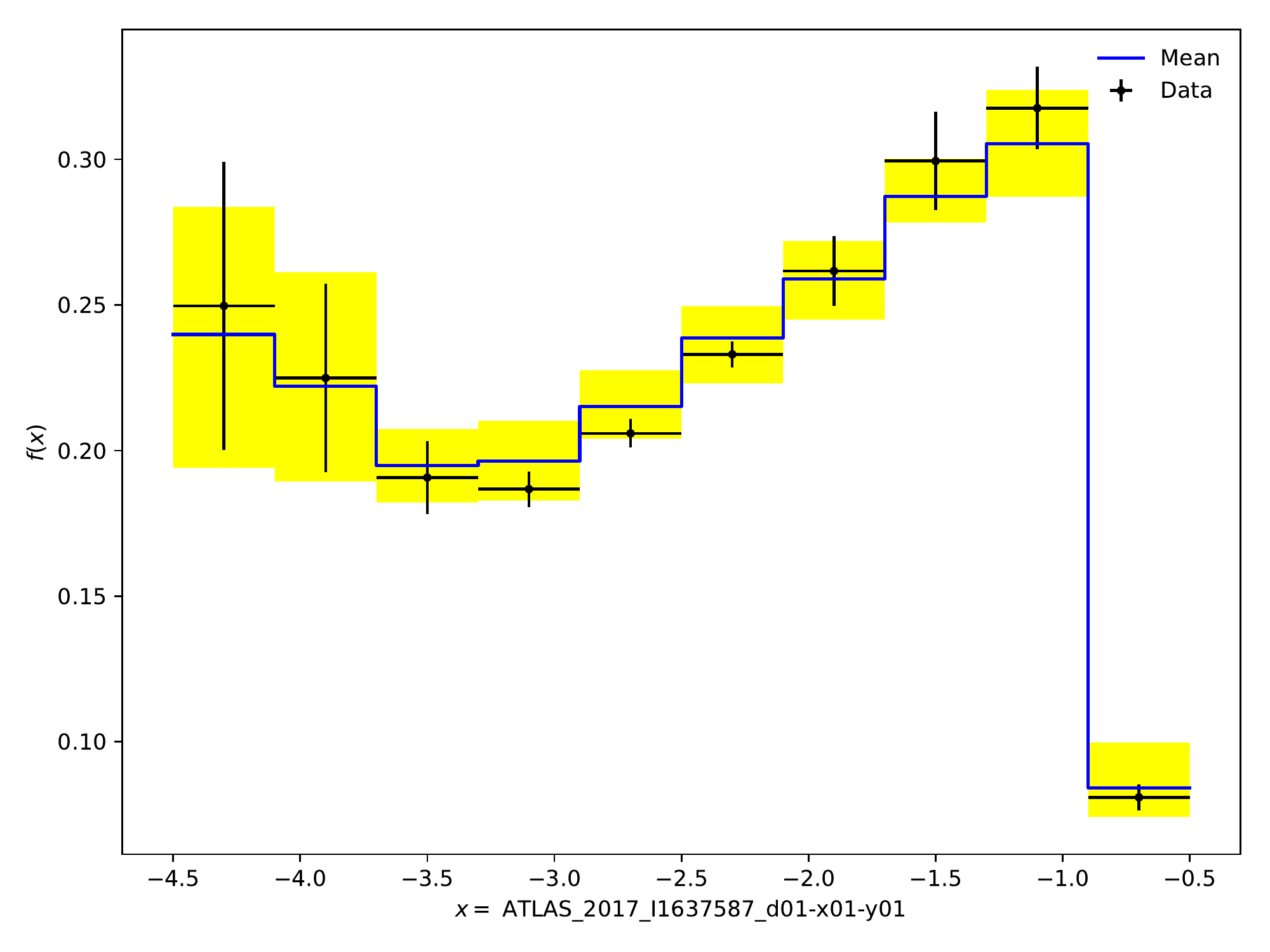}
         \includegraphics[width=0.47\textwidth]{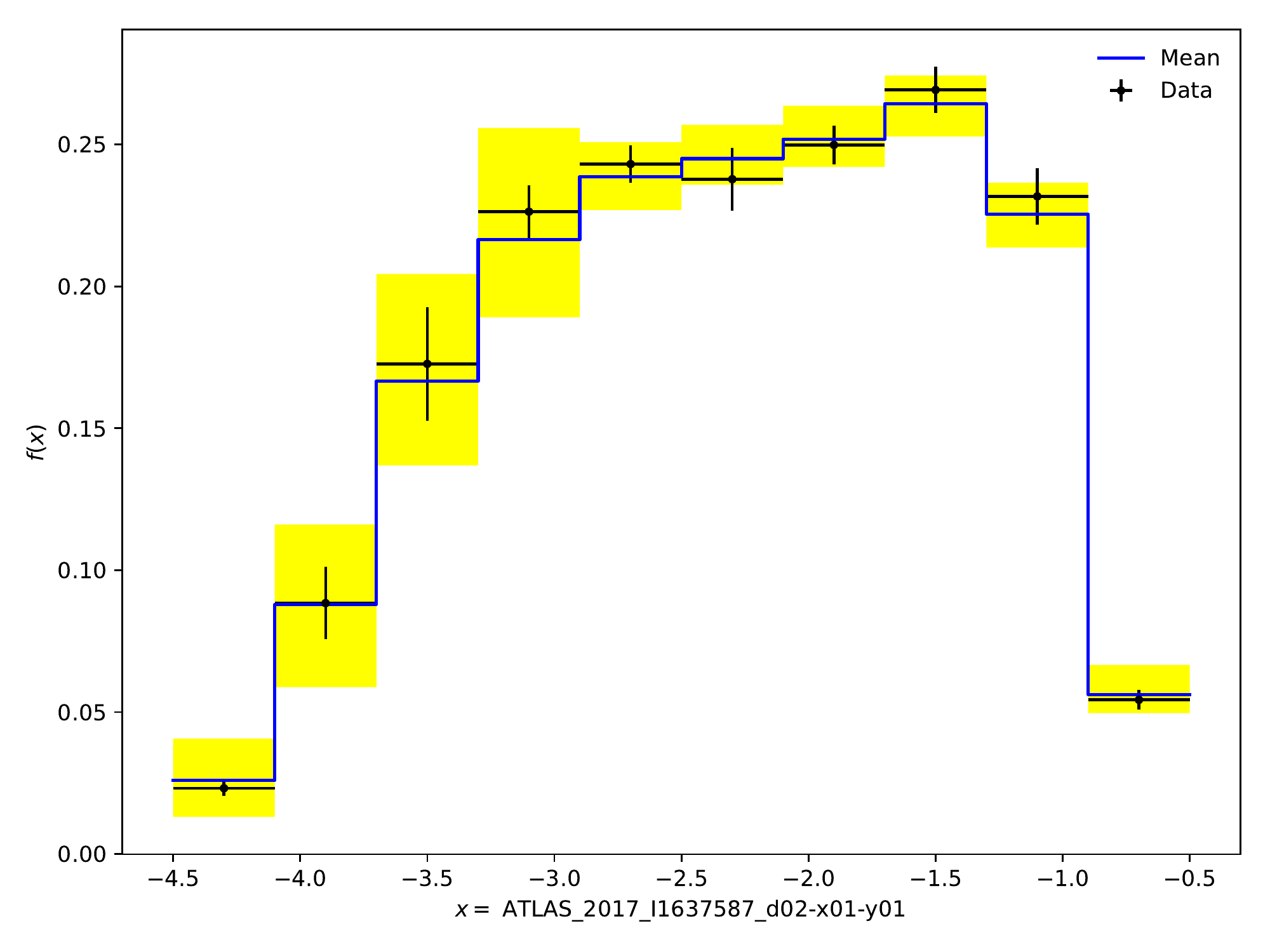}
             \centering
         \includegraphics[width=0.47\textwidth]{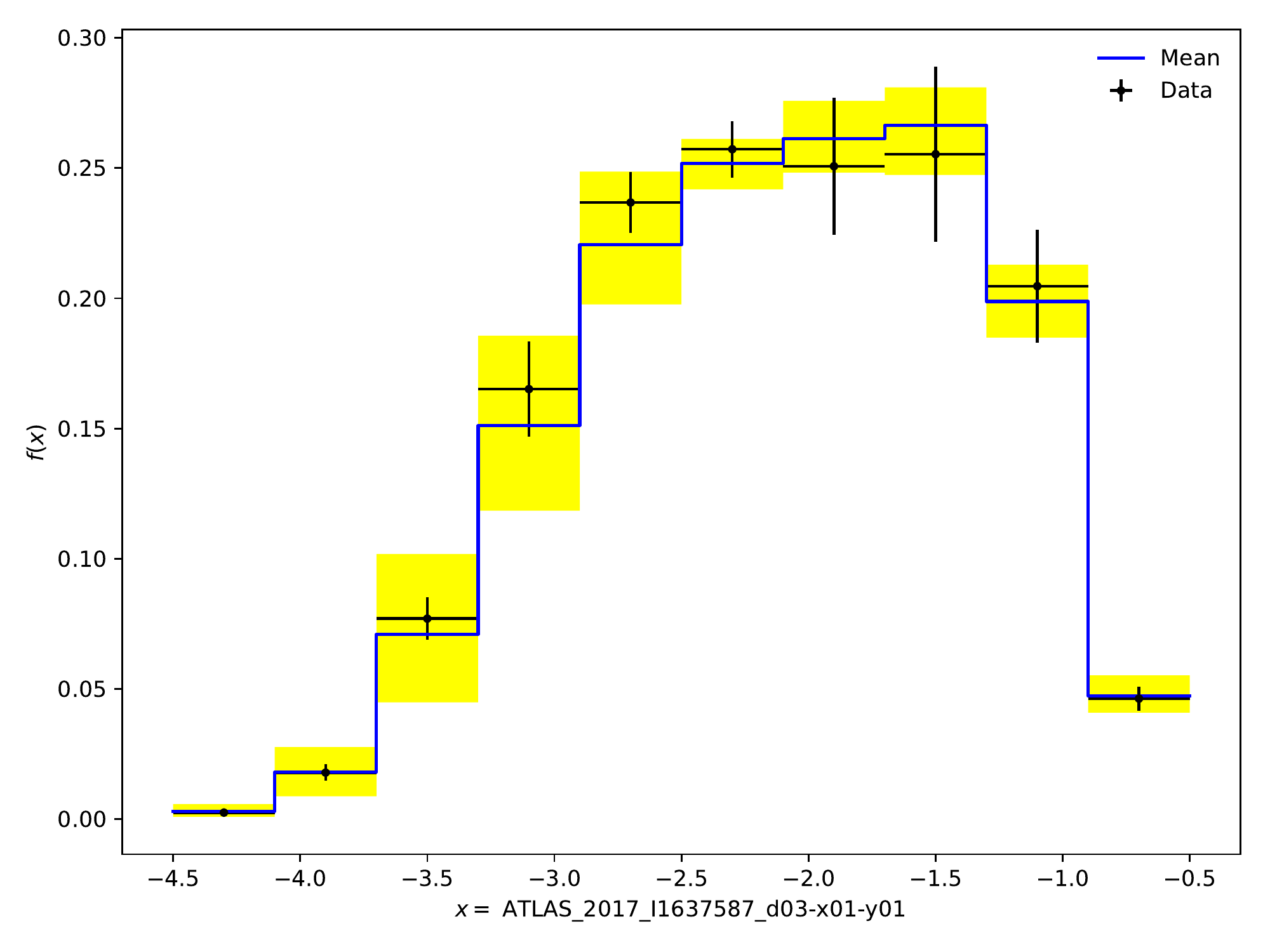}
     \caption{Envelope plots for ATLAS\_2017\_I1637587}
    \label{fig:ESDM1}
    \end{figure}

As can be seen in Figure \ref{fig:ESDM1}, The envelopes cover the reference data in almost every bin and hence we can say that the range selected for the parameters are appropriate. 

\FloatBarrier

\subsubsection{Lund Jet Plane Distributions}

     \begin{figure}[ht]
     \centering
         \includegraphics[width=0.47\textwidth]{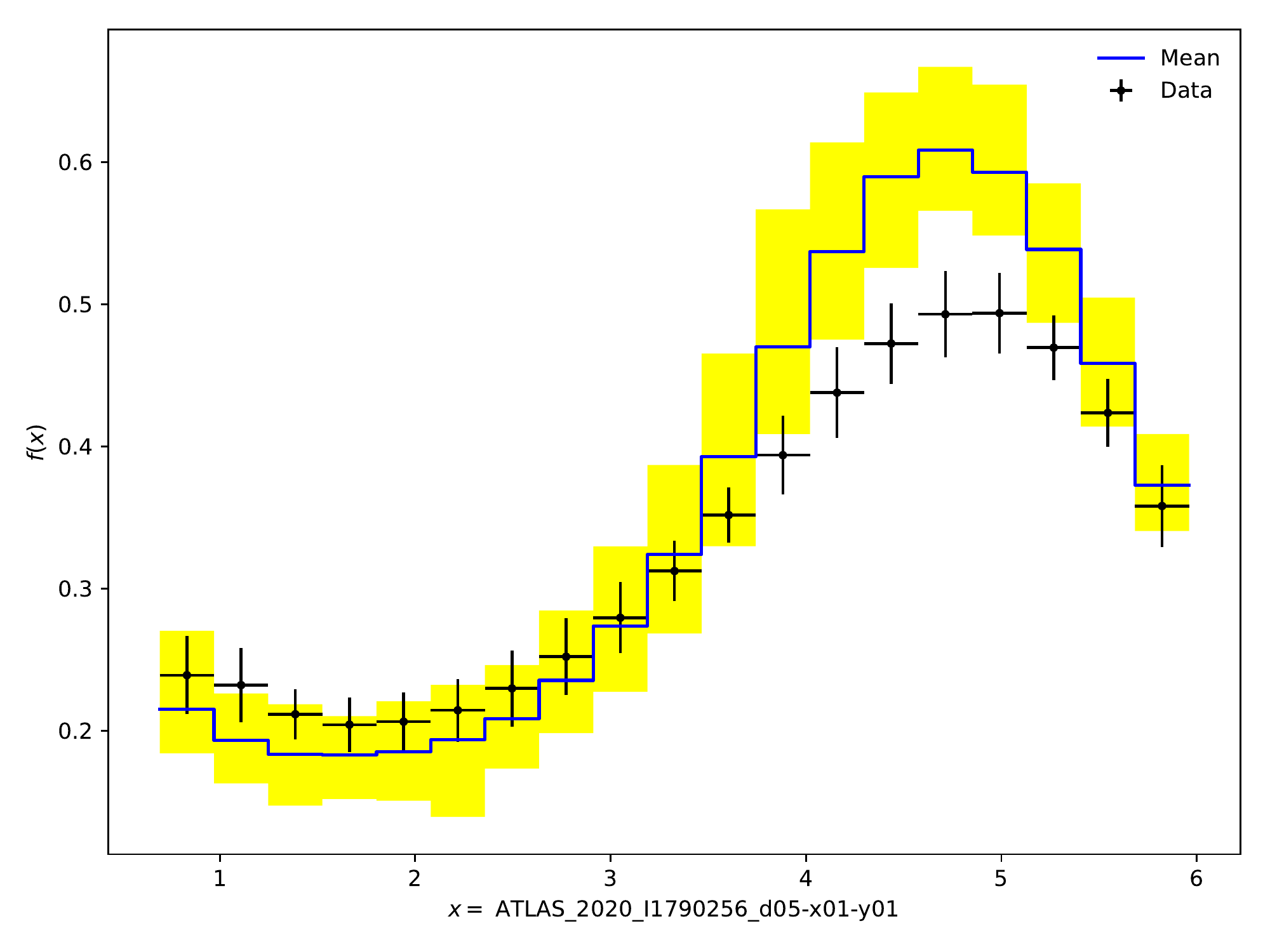}
         \includegraphics[width=0.47\textwidth]{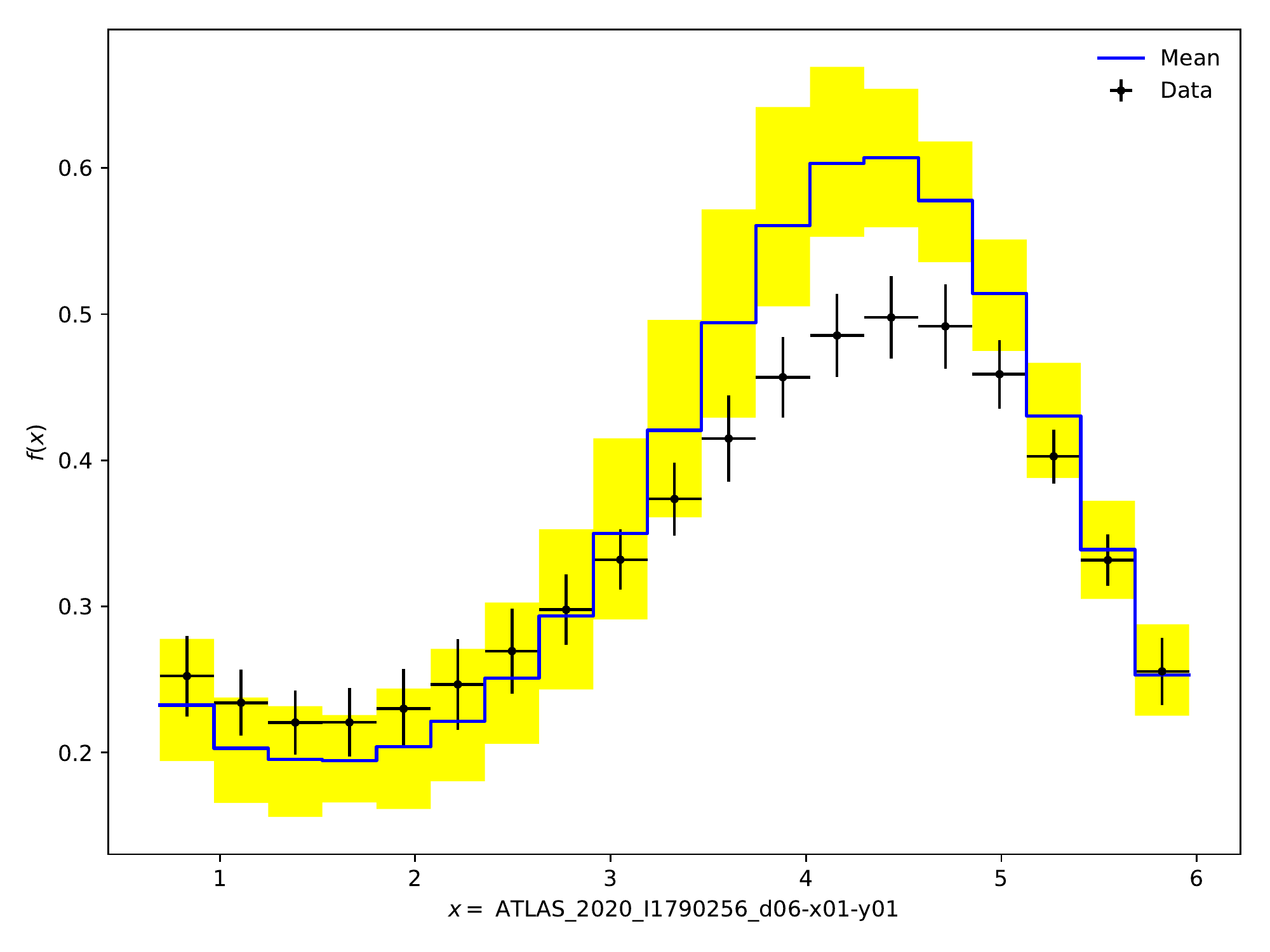}
         \includegraphics[width=0.47\textwidth]{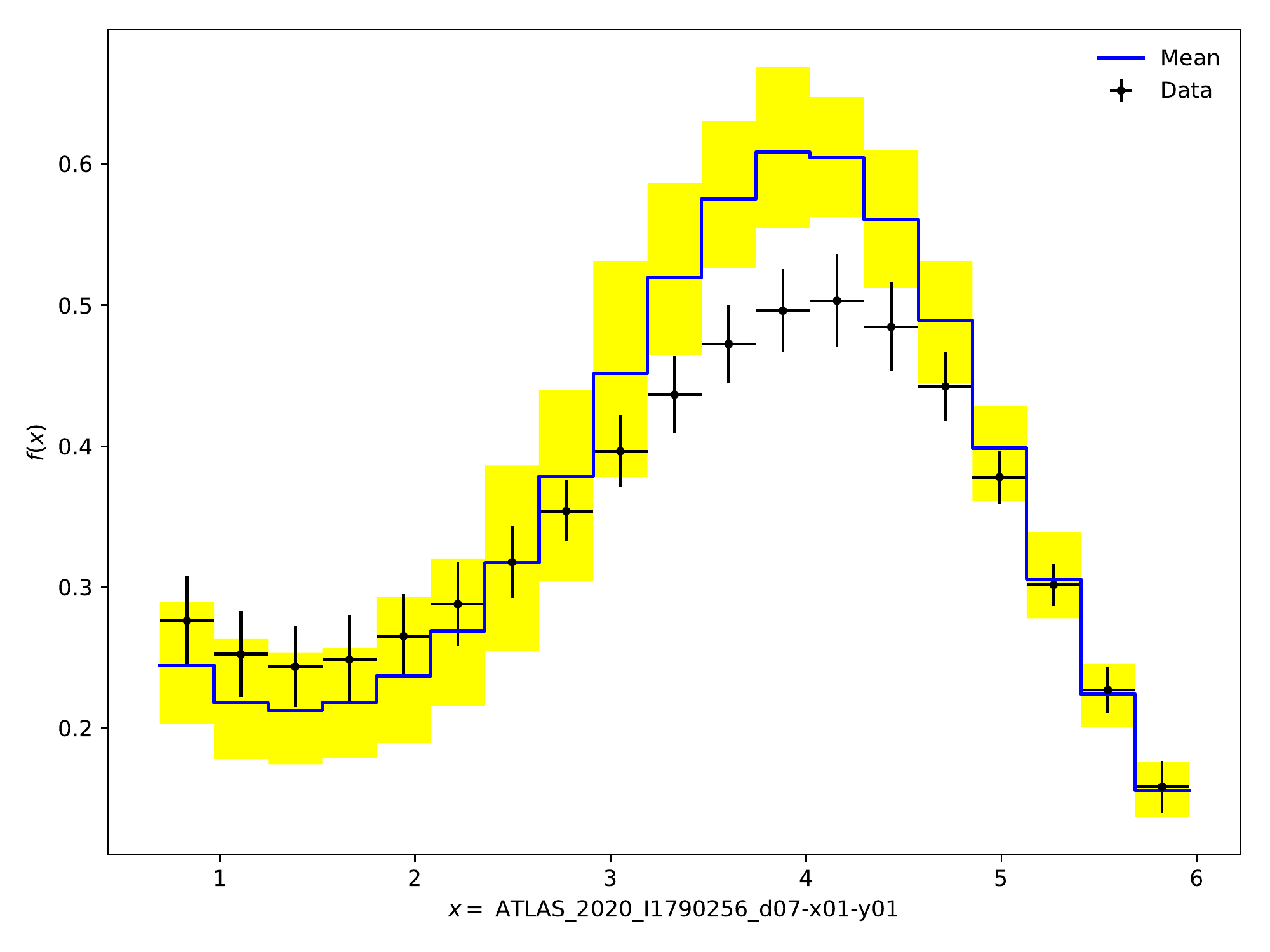}
         \includegraphics[width=0.47\textwidth]{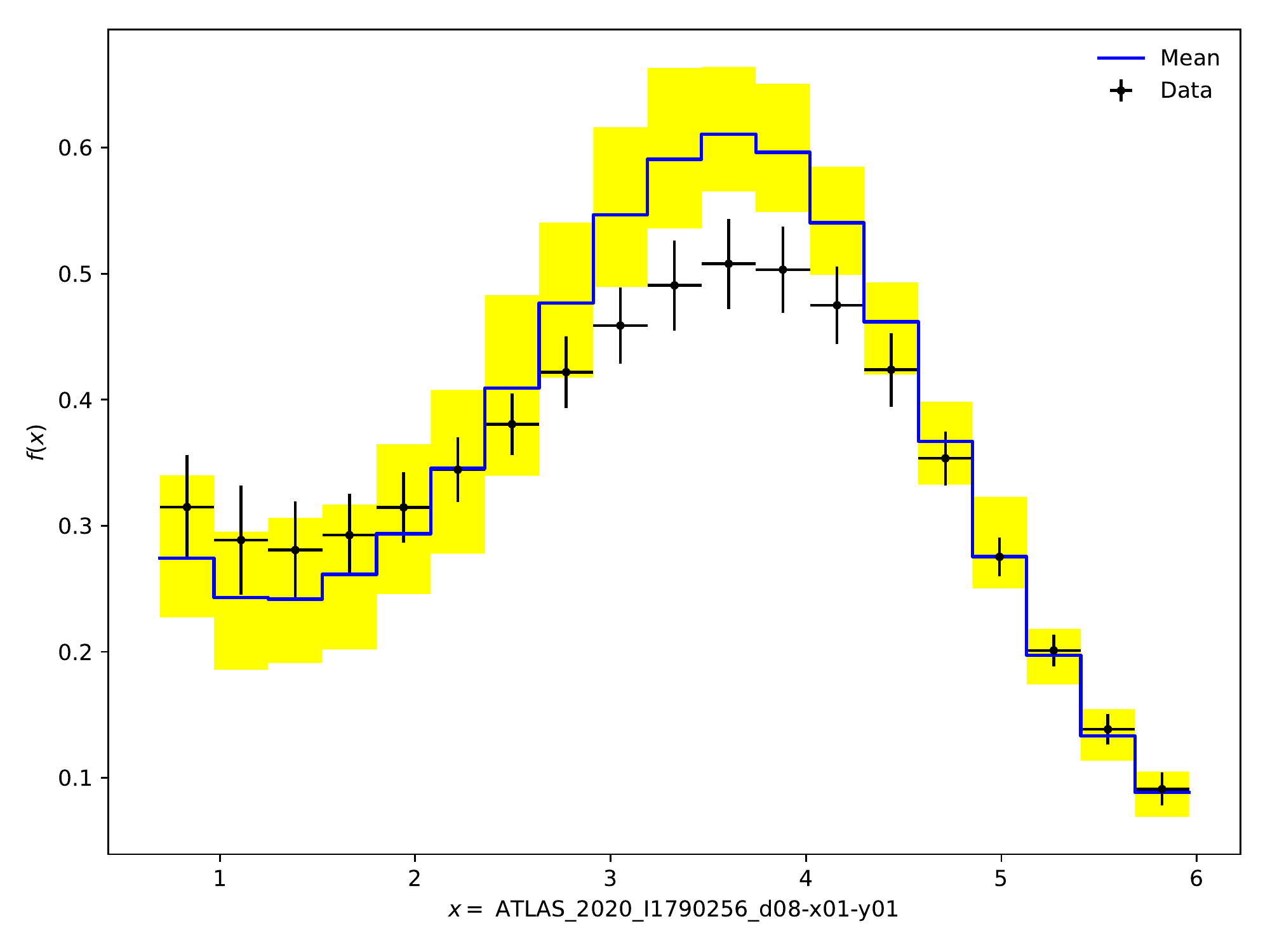}
    \caption{Envelope plots for Lund Jet Plane}
    \label{fig:ELJP1}
    \end{figure}

    As can be seen in Figure \ref{fig:ELJP1}, the envelopes do not entirely cover the reference data. This is because we have reached a limit as to how much the distribution can be further fitted to the data with Pythia. Thus we consider this suitable for the purpose of this report and proceed with our set of parameter ranges.
    
\FloatBarrier

\subsubsection{Soft Drop Observables Distributions}

\begin{figure}[ht]
     \centering
         \includegraphics[width=0.47\textwidth]{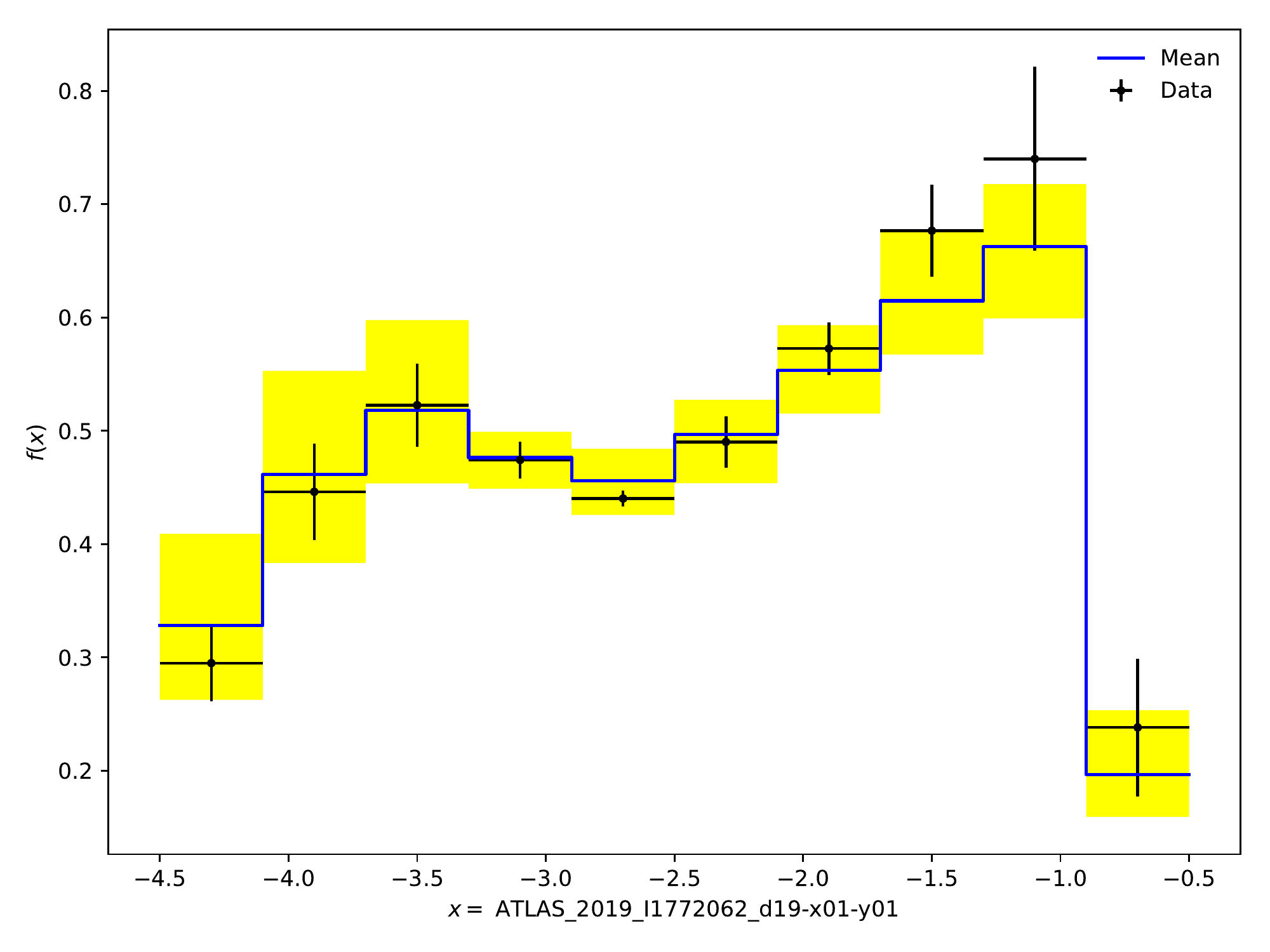}
         \includegraphics[width=0.47\textwidth]{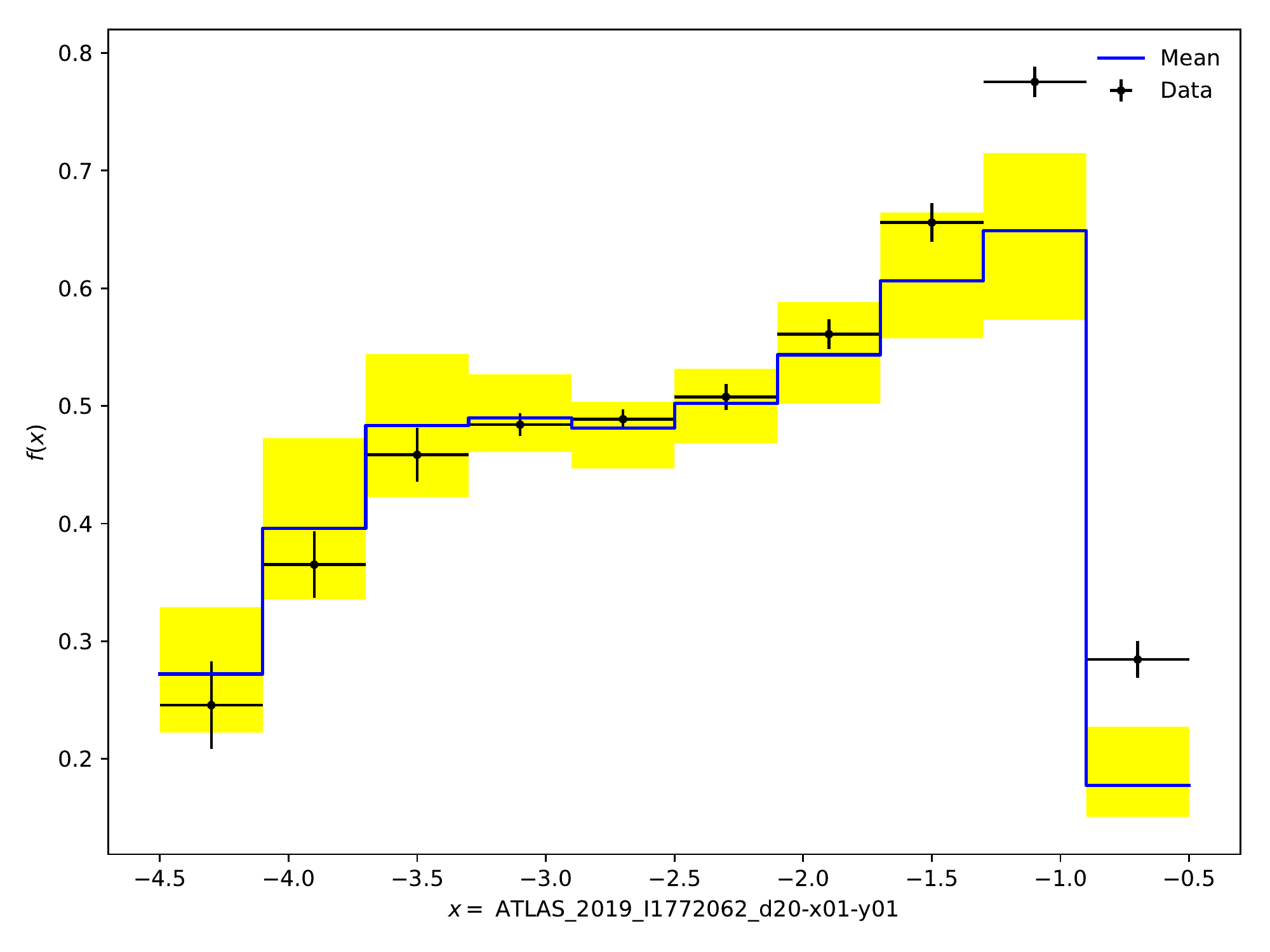}
         \includegraphics[width=0.47\textwidth]{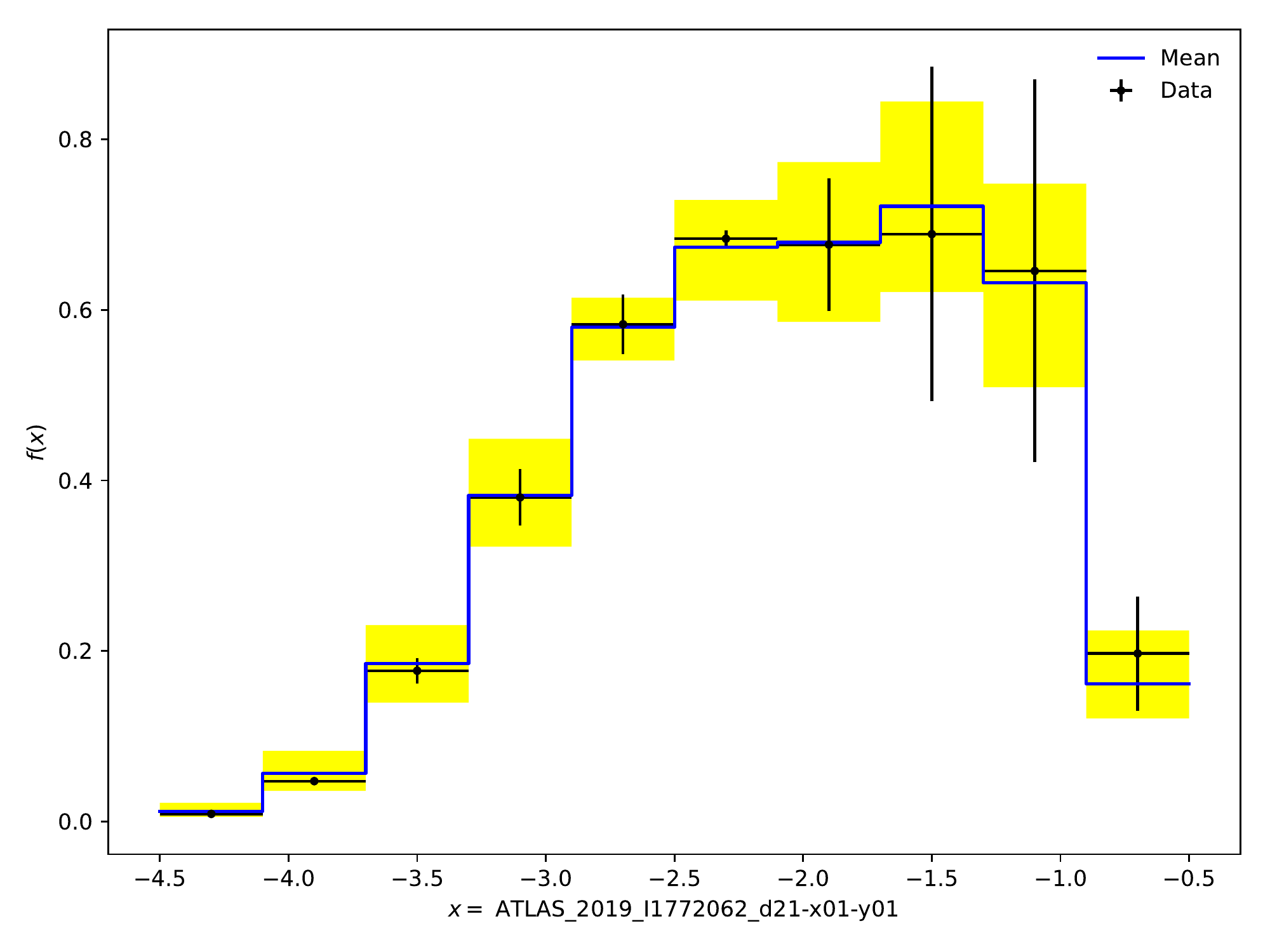}
         \includegraphics[width=0.47\textwidth]{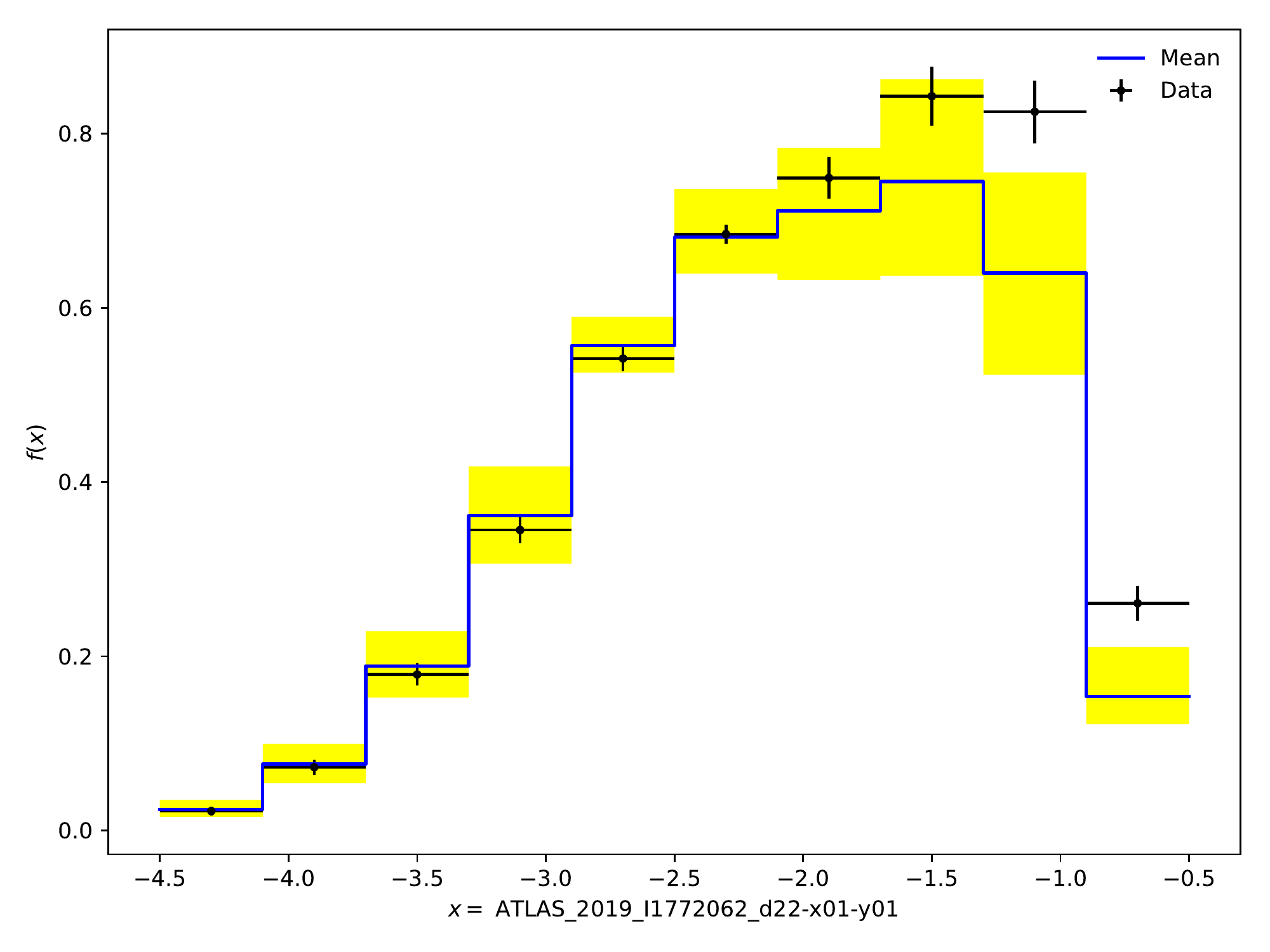}
         \includegraphics[width=0.47\textwidth]{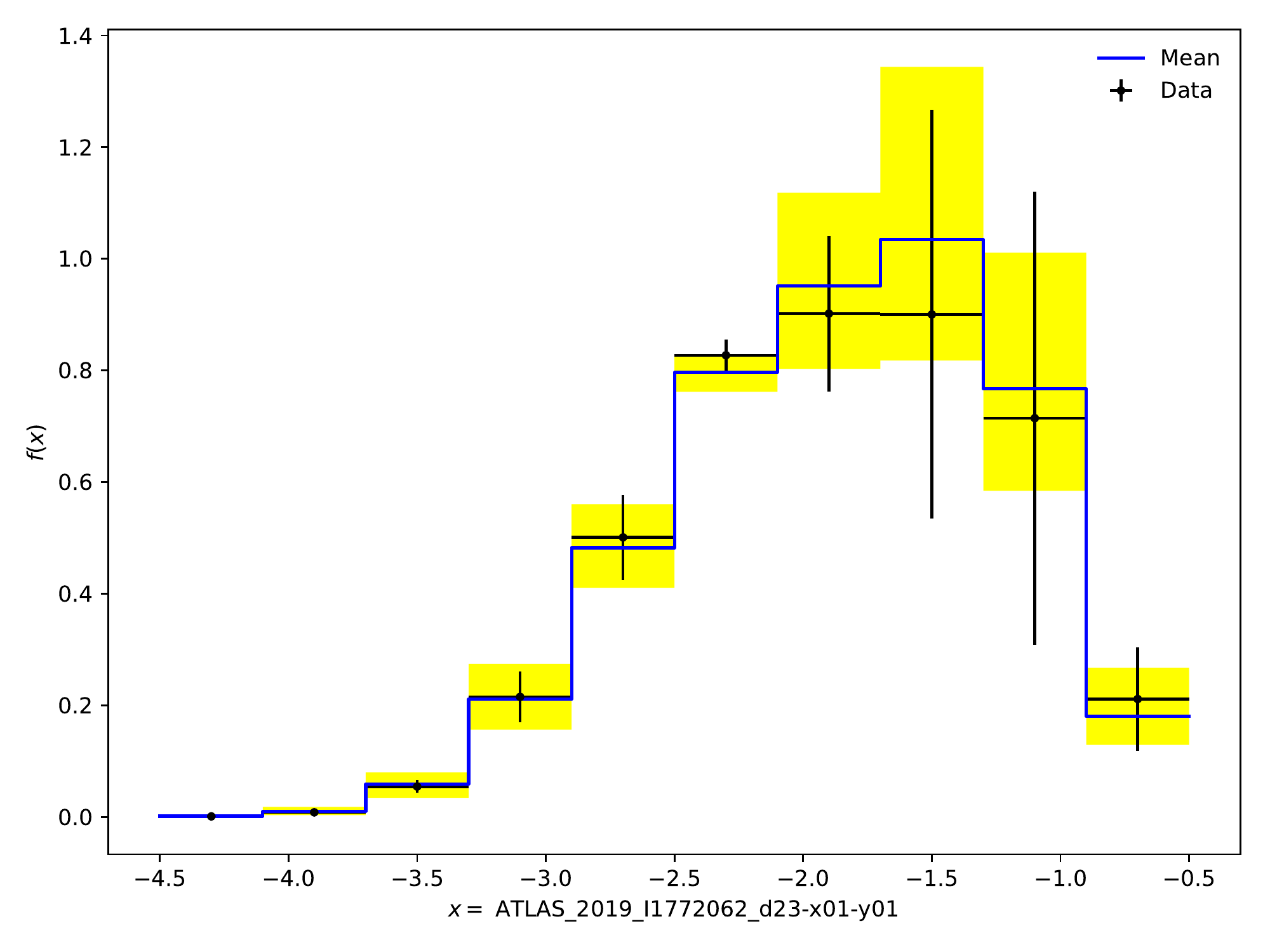}
         \includegraphics[width=0.47\textwidth]{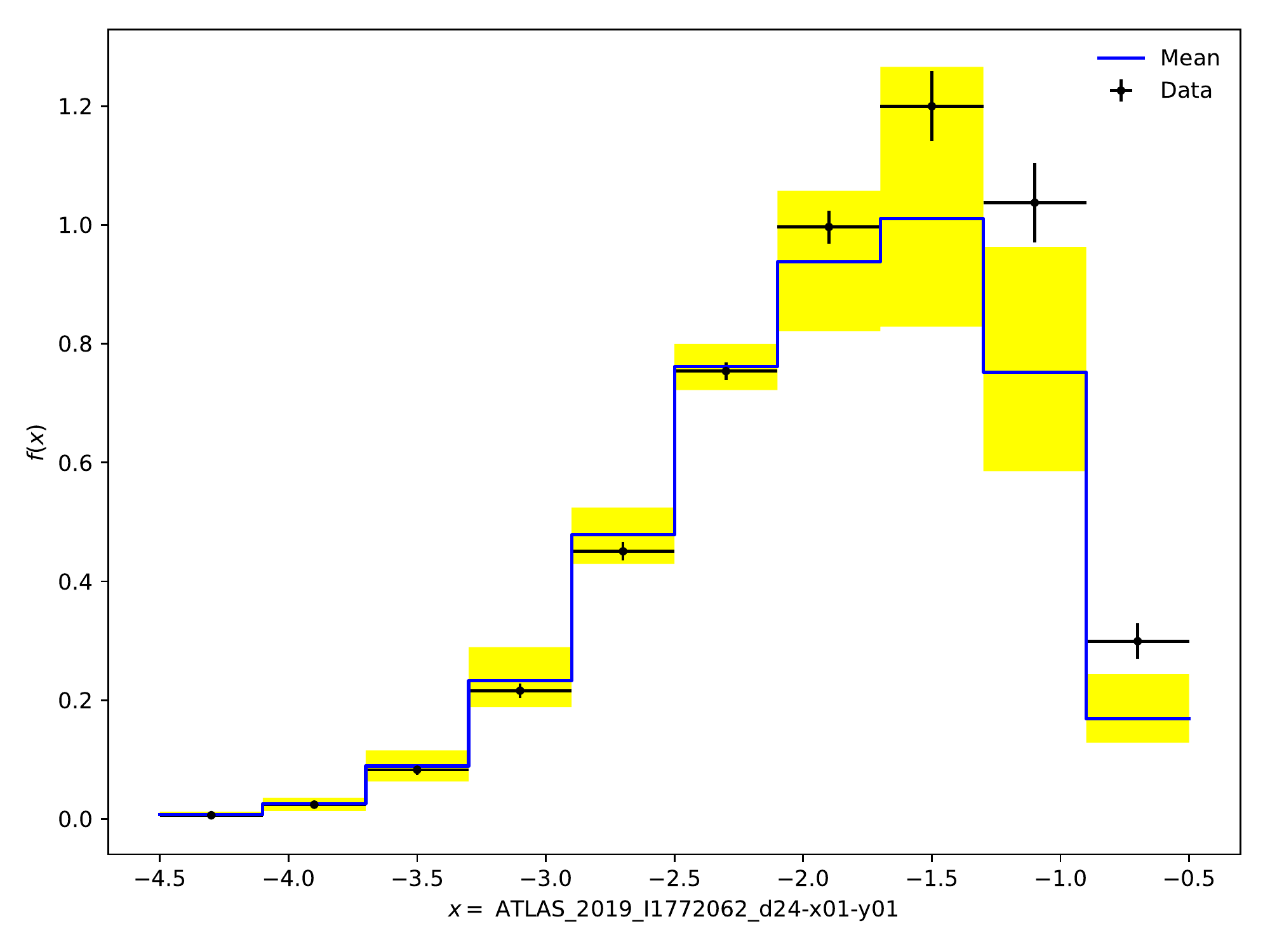}
    \caption{Envelope plots for Soft Drop Mass observable}
    \label{fig:Esd}
    \end{figure}

In the Figure \ref{fig:Esd}, we see that the envelopes cover the data points in almost all the bins of our distributions of interest i.e soft drop jet mass from the soft drop jet observables analysis. Thus the parameter ranges are suitable for proceeding to tune the distributions. 

\FloatBarrier

\subsection{Weight file}
\label{sec:wt}

The weight file assigns weights to the distributions to be tuned and hence is manually changed depending on our interests. For this project, a total of 16 distributions were assigned weights greater than 1 and 6 distributions were given no weight i.e 0 for obtaining the Common Tune. 
These are :

 \begin{table}[ht]
        \small
        \centering
        \begin{tabular}{ccc}
        Analysis & Distribution code & Weight\\
        \hline
        ATLAS\_2020\_I1790256 & d03-x01-y01 & 106\\
        ATLAS\_2020\_I1790256 & d04-x01-y01 & 112\\
        ATLAS\_2020\_I1790256 & d05-x01-y01 & 108\\
        ATLAS\_2020\_I1790256 & d06-x01-y01 & 20\\
        ATLAS\_2020\_I1790256 & d07-x01-y01 & 16.6\\
        ATLAS\_2020\_I1790256 & d08-x01-y01 & 16\\
        ATLAS\_2020\_I1790256 & d09-x01-y01 & 16\\
        ATLAS\_2019\_I1772062 & d19-x01-y01 & 75\\
        ATLAS\_2019\_I1772062 & d20-x01-y01 & 75\\
        ATLAS\_2019\_I1772062 & d21-x01-y01 & 200\\
        ATLAS\_2019\_I1772062 & d22-x01-y01 & 80\\
        ATLAS\_2019\_I1772062 & d23-x01-y01 & 200\\
        ATLAS\_2019\_I1772062 & d24-x01-y01 & 80\\
        ATLAS\_2017\_I1637587 & d01-x01-y01 & 500\\
        ATLAS\_2017\_I1637587 & d02-x01-y01 & 500\\
        ATLAS\_2017\_I1637587 & d03-x01-y01 & 500\\
        ATLAS\_2019\_I1772062 & d61-x01-y01 & 0\\
        ATLAS\_2019\_I1772062 & d62-x01-y01 & 0\\
        ATLAS\_2019\_I1772062 & d79-x01-y01 & 0\\
        ATLAS\_2019\_I1772062 & d80-x01-y01 & 0\\
        ATLAS\_2019\_I1772062 & d97-x01-y01 & 0\\
        ATLAS\_2019\_I1772062 & d98-x01-y01 & 0\\
      \end{tabular}
        \caption{Weights assigned to obtain the Common tune}
        \label{tab:weight}
    \end{table}

\FloatBarrier

\end{document}